\documentclass[aps,prl,twocolumn,showpacs,amsmath,amssymb,superscriptaddress]{revtex4-1}
\usepackage{graphicx}
\usepackage{color}
\usepackage{accents}
\usepackage{hyperref}
\hypersetup{
    colorlinks=true,
    linkcolor=blue,
    filecolor=blue,      
    urlcolor=blue,
    citecolor=blue
}

\usepackage[normalem]{ulem}
\usepackage{subfigure}

\def\be{\begin{equation}}
\def\ee{\end{equation}}
\def\ba{\begin{eqnarray}}
\def\ea{\end{eqnarray}}
\def\bc{\begin{center}}
\def\ec{\end{center}}

\renewcommand{\vec}[1]{\mathbf{#1}}

\begin{document}
\title{Back action in quantum electro-optic sampling of electromagnetic vacuum fluctuations}

\author{T. L. M. Guedes}

\affiliation{Department of Physics and Center for Applied
Photonics, University of Konstanz, D-78457 Konstanz, Germany}

\author{I. Vakulchyk}

\affiliation{Center for Theoretical Physics of Complex Systems, Institute for Basic Science (IBS), Daejeon 34126, Republic of Korea}

\affiliation{Basic Science Program, Korea University of Science and Technology (UST), Daejeon, Korea, 34113}

\author{D. V. Seletskiy}

\affiliation{Department of Engineering Physics, Polytechnique Montr\'{e}al, Montr\'{e}al, QC, H3T 1J4, Canada}

\author{A. Leitenstorfer }

\affiliation{Department of Physics and Center for Applied Photonics, University of Konstanz, D-78457 Konstanz, Germany}

\author{A. S. Moskalenko}

\email{moskalenko@kaist.ac.kr }

\affiliation{Department of Physics, KAIST, Daejeon 34141, Republic of Korea}

\author{Guido Burkard}

\email{guido.burkard@uni-konstanz.de}

\affiliation{Department of Physics and Center for Applied Photonics, University of Konstanz, D-78457 Konstanz, Germany}

\begin{abstract}

The influence of measurement back action on electro-optic sampling of electromagnetic quantum fluctuations is investigated. Based on a cascaded treatment of the nonlinear interaction between a near-infrared coherent probe and the mid-infrared vacuum, we account for the generated electric-field contributions that lead to detectable back action. Specifically, we theoretically address two realistic setups, exploiting one or two probe beams for the nonlinear interaction with the quantum vacuum, respectively. The setup parameters at which back action starts to considerably contaminate the measured noise profiles are determined. Due to the vacuum fluctuations entering at the beam splitter, the shot noise of two incoming probe pulses in different channels is uncorrelated. This leads to the absence of the base-level shot noise in the correlation, while further contributions due to nonlinear shot-noise enhancement are still present. Ultimately, the regime in which  electro-optic sampling of quantum fields can be considered as effectively back-action free is found.

\end{abstract}

\maketitle

The advent of quantum mechanics has revolutionized physics and also deeply influenced several other branches of science, reaching as far as quantum biology~\cite{quantumbiology}, biochemistry~\cite{biochem}, quantum spectroscopy \cite{Schlawin2017}, and quantum information science~\cite{qubit}. The uncertainty relations belong to the most remarkable features of quantum theory that follow directly from the fundamental quantization rules. Non-commuting observables cannot be simultaneously determined with arbitrarily high precision, since the product of their uncertainties has a lower bound imposed by nature itself~\cite{Heisenberg,Robertson1929,Werner2019}. 

When performing a quantum measurement, the interaction with the measurement device typically leads to a perturbation of the  state of the probed system. Even for experiments keeping the product of related uncertainties at their minimum, improving the accuracy with which a given observable is measured inevitably increases the fluctuations in its (non-commuting) canonically conjugate variable, a clear demonstration of how a measurement can affect a quantum system. This kind of influence of the measurement device on a quantum system is called quantum back action (BA)~\cite{braginsky}. 
Often, the BA is undesired, but in some cases, e.g., for the purpose of quantum error correction~\cite{Terhal2015} or measurement-based quantum computation~\cite{Briegel2009}, it underlies the functionality of quantum-information processing schemes.

Fluctuations in non-commuting observables persist even when the system reaches its lowest possible energy content, its ground state, a feature known as zero-point fluctuations. In recent years, several remarkable experiments have been carried out aiming at probing the zero-point fluctuations of a plethora of quantum systems, in particular single-mode mesoscopic mechanical resonators~\cite{nanoresonator} and multi-mode electromagnetic radiation~\cite{vacuum_samp, subcycle, benea_vac}. 

Theoretical and experimental evidence~\cite{khalili, spethmann} points towards the inevitable presence of BA in experiments involving quantum mechanical resonators probed by light in optical cavities. While the light affects the resonator through radiation pressure (or analogously through Stokes and anti-Stokes scattering processes), the resonator imprints its phase-space signature on the photons in the cavity, or, correspondingly, shifts the resonance frequency of the cavity~\cite{optomechanicsreview}. This BA, however, can be avoided by coupling the vibrational modes of two oscillators through the cavity photons, allowing the BA contributions from the two modes to cancel each other~\cite{rodrigo}.

Related arguments based on mode coupling through BA have been invoked to explain the results of electro-optic (EO) measurements of correlations in the vacuum state of the electromagnetic field~\cite{benea}. The potential effect of BA in such experiments, however, might considerably diverge from those seen in optomechanical cavities~\cite{chaos, nonlinear+qubit}, since the characteristic nonlinearity of the EO interaction effectively couples optical modes between and within channels, each of which consisting of an infinite and continuous set of modes.

\begin{figure}
    \centering
    \includegraphics[width=\linewidth]{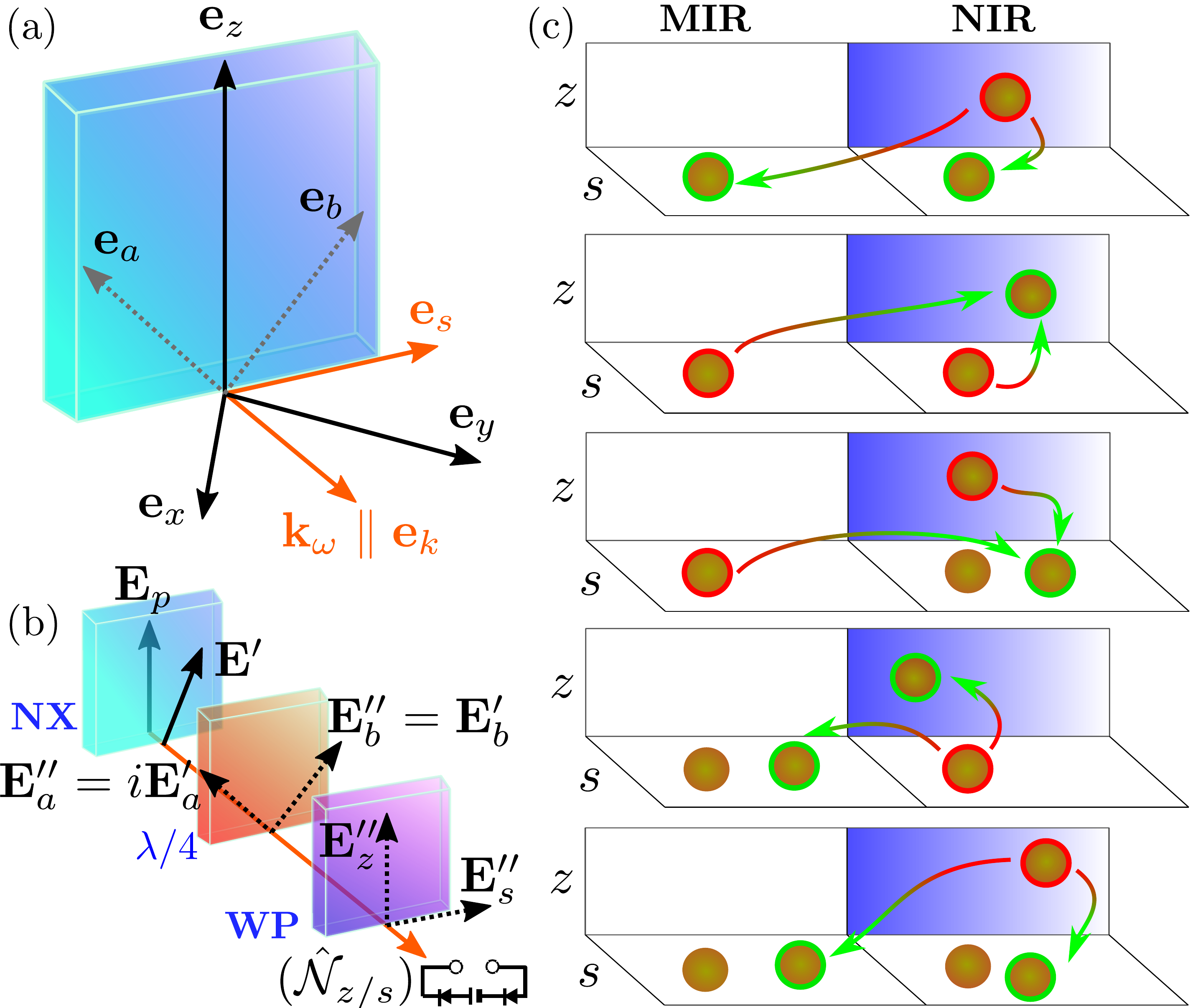}
 \caption{(a) Orientation of the NX and field components. The probe $\vec{E}_\mathrm{p}$ is polarized along $\vec{e}_z$ and propagates with wave vector $\vec{k}_\omega = k_\omega \vec{e}_k$. The tensor components of the nonlinear susceptibility are such that only field components along $\vec{e}_x$ and $\vec{e}_y$ can mix with the probe, generating new quantum-field components along $\vec{e}_s$ and $\vec{e}_k$ (a nonpropagating component). (b) Scheme of an EO measurement. $\vec{E}_\mathrm{p}$ changes its ellipticity in the NX due to the nonlinearly generated $\vec{e}_s$-polarized field component, resulting in $\vec{E}'$. A quarter-wave plate ($\lambda/4$) shifts the phase of the $\vec{e}_a$ component of $\vec{E}'$ by $\pi/2$, leading to $\vec{E}''$, which has its ${E}''_z$ and ${E}''_s$ components spatially split by a Wollaston prism (WP) before independent photon counting. (c) Illustration of the state evolution under the nonlinear interaction~\cite{Suppl_Mat}. The states include $z$ and $s$ polarizations (in-plane and out-of-plane panels) and two frequency bands, MIR and NIR (left and right panels), with the $z$-polarized NIR coherent probe represented as a blue pool of photons. Red (green) contours represent annihilation (creation) of photons (golden spheres) and the arrows show the directions of the energy transfers. The 1st diagram (uppermost) shows the lowest-order perturbation of the initial state, with $s$-polarized MIR and NIR photons being created through annihilation of a probe photon. The 2nd and 3rd diagrams show the 2nd-order processes that lead to no BA in the MIR, since the photons created by the 1st-order process are annihilated. The last two diagrams show the remaining 2nd-order processes, which cause additional BA in the MIR via generation of extra photons.}
  \label{fig1}
\end{figure}

In this Letter, we theoretically study the BA in two experimental settings involving EO sampling of the electromagnetic ground state~\cite{benea, vacuum_samp}. For a single-channel experiment~\cite{vacuum_samp} [see Fig.~\ref{fig1}(b)], the interplay between shot noise (SN) and BA plays a crucial role in determining the optimal range of parameters. The situation changes considerably when a second channel is included in the setup~\cite{benea, lightcone_correlation} [as in Fig.~\ref{autocorrelation}(a)], allowing for evasion of the base-SN contribution to the signal correlation. We propose the working regimes most suitable to avoid major BA contributions to the EO signals and explain the subtle role that the population of the measured modes plays in the data presented in Ref.~\cite{benea}. 

We start with a derivation of the cascaded contributions to the quantum electric field that build up within the nonlinear medium and proceed to their conversion into the EO signal noise upon detection.
We initially consider a single-channel setup as sketched in Figs.~\ref{fig1}(a) and \ref{fig1}(b). An incoming near-infrared (NIR) ultrashort probe pulse copropagates with a mid-infrared (MIR) vacuum-state electric-field component along the $\vec{e}_k$=[110] axis of a zinc-blende-type nonlinear crystal (NX), as in Ref.~\cite{Moskalenko}. The wave vector $\vec{k}_\omega=k_\omega \vec{e}_k$ of the probe pulse is perpendicular to the $z$ axis of the crystal, which in turn is parallel to the probe electric field $\vec{E}_\mathrm{p}= E_\mathrm{p}\vec{e}_z$. Since the probe pulse is in a coherent state, we can write its field operator as $\hat{\vec{E}}_\mathrm{p} (\vec{r}, t) = E_\mathrm{p}(\vec{r}, t)\vec{e}_z+\delta \hat{\vec{E}} (\vec{r}, t)$, with $\delta \hat{\vec{E}} (\vec{r}, t)$ describing the (zero-point) quantum fluctuations. The (cascaded) contribution to the second-order nonlinear polarization arising in the crystal from the mixing between the probe and any $s$-polarized (with $\vec{e}_s=\vec{e}_z\times \vec{e}_k$) $m$th-order quantum-field contribution $\hat{E}^{(m)}_s$ present in the crystal is given by
$\hat{\vec{P}}^{(2)}_m (\vec{r}, t)=-\epsilon_0d \hat{E}^{(m)}_{s} (\vec{r}, t)E_\mathrm{p}(\vec{r}, t)\vec{e}_s$.
Here, $\epsilon_0$ is the vacuum permittivity and $d=-n^4r_{41}$ is the effective nonlinear susceptibility of the NX, with $n$ being its refractive index (RI) at the center frequency of the probe and $r_{41}$ its relevant EO coefficient~\cite{Boyd_book}. 
 
 We divide the frequency domain into two segments, the MIR, represented by $\Omega$, and the NIR, represented by $\omega$, both of which can be expressed by $\Lambda \in \{\Omega, \omega\}$. The electric-field operator is given in its paraxial form~\cite{Allen1992, Calvo2006} $\hat{\vec{E}}(\vec{r},t)= \sum_{\sigma l p} \hspace{-0.6mm}\int \hspace{-0.6mm} \mathrm{d}\Lambda  \vec{e}_{\sigma} 
 e^{i\Lambda (n_\Lambda r_k/c_0-t)} \hat{{E}}_{\sigma lp}(\vec{r},\Lambda)$, where
 \begin{equation}
\hat{{E}}_{\sigma lp}(\vec{r},\Lambda)\!=\!  i\text{sgn}(\Lambda)\!\sqrt{\frac{\hbar|\Lambda| }{4\pi\epsilon_0n_\Lambda c_0}}\mathrm{LG}_{lp} (\vec{r}_\perp, \Delta r_k; \Lambda)\hat{a}_{\sigma l p}( \Lambda), \label{EFielddecompositionparaxial3}
\end{equation}
 with amplitude distributions (dependent on $\vec{r}_\perp=r_z\vec{e}_z+r_s\vec{e}_s$) in the transversal plane located at $r_k=\vec{r}\cdot\vec{e}_k$ given by the Laguerre-Gaussian (LG) modes $\mathrm{LG}_{lp} (\vec{r}_\perp,\Delta r_k= r_k-L/2; \Lambda)$. The operator $\hat{a}_{\sigma lp}(\Lambda>0)=\hat{a}^\dagger_{\sigma lp}(-\Lambda)$~\cite{Calvo2006,me} annihilates a photon with frequency $\Lambda$, polarization $\sigma=s,z$, and azimuthal $l$ and radial $p$ indices labeling the LG modes. Here, $c_0$ and $n_\Lambda$ are the speed of light and the frequency-dependent RI of the NX,
respectively.
We assume an NX extended from $r_{k}=0$ to $r_{k}=L$, so that the narrowest amplitude distribution for the LG modes occurs at the center of the crystal, $r_k=L/2$, where the waist of the mode profiles reaches its minimal value $w_0$. $E_\mathrm{p}$ is given by Eq.~\eqref{EFielddecompositionparaxial3} with $\hat{a}_{\sigma l p}( \omega)\to\alpha_{\sigma l p}( \omega)$,
and we assume its transversal profile to be in the fundamental mode, given by $g_{_{00}}(\vec{r}_{\!{_\perp}})\!\equiv\!\mathrm{LG}_{_{00}}(\vec{r}_{\!{_\perp}},0;\omega)=\sqrt{2/\pi}\,w_0^{-1} \exp(-r_{\!{_\perp}}^2/w_0^2)$ and, accordingly, $\alpha_\mathrm{p}( \omega)\equiv\alpha_{z 0 0}(\omega)$. For the remainder of this work, we shall exploit the fact that the NX is thin  ({$L\ll n_\Lambda \Lambda w^2_0/2c_0$}, for the frequencies $\Lambda$ of interest) and consider all fields at $r_{k}=L/2$: $\hat{{E}}_{\sigma lp}(\{r_{k}=L/2,\vec{r}_{\!{_\perp}}\},\Lambda)\equiv\hat{{E}}_{\sigma lp}(\vec{r}_{\!{_\perp}},\Lambda)$. For brevity, the indices $l$ and $p$ will be omitted whenever possible.

The nonlinear polarization acts as a source for the generation of multimode squeezed electric-field components~\cite{me, our_nature, Sho} with polarization perpendicular to the probe.
Neglecting the depletion of the probe pulse, the NIR output field of interest $\hat{\vec{E}}'$ is given by the input probe pulse plus the perturbative solutions to the wave equation sourced by each of the cascaded $\vec{P}^{(2)}_m(\omega)$,
\begin{equation}
  \begin{split}
  \hat{E}_{s}^{(m+1)}(\vec{r}_{\!{_\perp}}\!,\omega)\!=\!\!
  \int_{-\infty}^\infty\!\!\!\!\!\!\mathrm{d}\Omega\ \hat{E}^{(m)}_s(\vec{r}_{\!{_\perp}}\!,\Omega)
   E_{\mathrm{p}}(\vec{r}_{\!{_\perp}},\omega\!-\!\Omega) \zeta_{\omega,\Omega}
   \end{split},
   \label{E_NIR}
\end{equation}
where $\hat{E}^{(1)}_s (\Omega )\equiv\delta \hat{E}_s(\Omega )$ represents the vacuum electric field in the MIR 
and $\hat{E}^{(m)}_s(\Omega )$ for $m>1$ reads
\begin{equation}
  \begin{split}
  {\hat{E}}^{(m)}_s\!(\vec{r}_{\!{_\perp}}\!,\Omega)\!=\!\!
  \int_{-\infty}^\infty\!\!\!\!\!\!\!\mathrm{d}\omega  \hat{E}^{(m-1)}_s(\vec{r}_{\!{_\perp}}\!,\omega)
   E^*_{\mathrm{p}}(\vec{r}_{\!{_\perp}},\omega\!-\!\Omega) \zeta^*_{-\Omega,\Omega}
   \end{split}.
   \label{E_MIR}
\end{equation}
As can be seen from Eq.~\eqref{E_MIR}, the cascaded-field generation requires $\hat{E}^{(m-1)}_s (\omega )$ as given by Eq.~\eqref{E_NIR} to describe the higher-order MIR contributions {[in the same way that Eq.~\eqref{E_NIR} requires \eqref{E_MIR}]}, with  $\hat{E}^{(1)}_s (\omega )\equiv \delta \hat{E}_s(\omega )$ being the vacuum NIR electric field.
The factor $\zeta_{\pm\Lambda,\Omega}=\mp id\frac{L \Lambda}{2c_0 n }
   \mathrm{sinc}\!\left[\frac{L\Omega}{2c_0}(n_{_\Omega}-n_\mathrm{g})\right]\mathrm{exp}\left[i\frac{L\Omega}{2c_0}(n_{_\Omega}-n_\mathrm{g})\right]$
determines phase matching. Here 
$n_\mathrm{g}=c_0\partial k_\omega/\partial \omega$ is the group RI, taken at the central probe frequency.
The total generated field in each of the frequency ranges is then $\Delta\hat{\vec{E}}'(\Lambda)=\sum_{m>1} \hat{\vec{E}}^{(m)}(\Lambda)$, and accordingly $\hat{\vec{E}}'(\omega)=\vec{E}_\mathrm{p}(\omega)+\delta\hat{\vec{E}}(\omega)+\Delta\hat{\vec{E}}'(\omega)$ for the analyzed NIR.

After the NX, the detection part of the setup consists of an ellipsometer including two balanced photocounters that record the statistics of the photon numbers $ \hat{\mathcal{N}}_s$ and $ \hat{\mathcal{N}}_z$ for the $s$- and $z$-polarization components of the output NIR field  [cf. Fig.~\ref{fig1}(b)]. For the evaluation of the quantum signal,  $\hat{\mathcal{S}}=\hat{\mathcal{N}}_s-\hat{\mathcal{N}}_z$,
 we may neglect quadratic or higher-order terms in $\delta \hat{E}$.
  $\delta\hat{{E}}_s(\Lambda)$ is given by Eq.~\eqref{EFielddecompositionparaxial3} with transverse mode functions $g'_{_{lp}}(\vec{r}_{\!{_\perp}})\equiv\mathrm{LG}_{_{lp}}(\vec{r}_{\!{_\perp}},0;\Lambda)$. The total signal can be split into EO and base-SN contributions, $\hat{\mathcal{S}}=\hat{\mathcal{S}}_\mathrm{eo}+\hat{\mathcal{S}}_\mathrm{sn}$. In a perturbative approach, the contributions to the EO signal, $\hat{\mathcal{S}}_\mathrm{eo}=\sum^\infty_{j=1}\hat{\mathcal{S}}^{(j)}$, are given by 
\begin{equation}
  \hat{\mathcal{S}}^{(j)}=i\sqrt{B}{A^{(j+1)}}
   \int_{0}^\infty\!\!\mathrm{d}\Omega\, \sqrt{\frac{\Omega}{n_{_\Omega}}}\big[\hat{a}^{(j)}_{s00}(\Omega) R(\Omega)
  -\mathrm{H.c.}\big],
  \label{S_eo}
\end{equation}
in which ${A^{(j+1)}=\int\mathrm{d}^2 r_{_{\!\perp}}g_{_{00}}^{j+1}(\vec{r}_{_\perp})}
g'_{_{00}}(\vec{r}_{_\perp})$ and $\hat{a}^{(j)}_{s00}$ is the Bogoliubov-transformed (outgoing) annihilation operator, given by a series of nested convolutions of $\hat{a}_{s00}(\Lambda)$ and $\hat{a}^\dagger_{s00}(\Lambda)$ with functions covering either MIR or NIR frequencies depending on the value of $j$~\cite{our_nature, me}.
In Eq.~\eqref{S_eo},
$B={(d^2L^2N^2\omega_\mathrm{p}^2\hslash)\big/}{(4\pi\epsilon_0c_0^3n^2)}$ and 
$1/\omega_\mathrm{p}=\beta/\kappa$ is the average inverse detected frequency, with $\beta=\int_0^\infty\!\frac{\mathrm{d}\omega}{\omega}|\alpha_\mathrm{p}(\omega)|^2$ and $\kappa=\int_0^\infty\!\mathrm{d}\omega |\alpha_\mathrm{p}(\omega)|^2$.
We have introduced the expectation value of the photon number per probe pulse $N=\langle \hat{\mathcal{N}}_s+\hat{\mathcal{N}}_z \rangle=\frac{4\pi c_0 n \epsilon_0}{\hslash}\beta$ and the gating function
  $R(\Omega)=i\zeta_{\omega,\Omega}F(\Omega)/(d\frac{L \omega}{2c_0 n })$ 
  with $F(\Omega)=\frac{1}{2}[{f}^*_+(\Omega)+{f}_-(\Omega)]$,
and $f_\pm(\Omega)=\int_0^\infty\!\!\mathrm{d}\omega
{\alpha}_\mathrm{p}^*(\omega){\alpha}_\mathrm{p}(\omega\pm\Omega)/\kappa$.
The base-SN contribution is given by
\begin{equation}
\hat{\mathcal{S}}_{\text{sn}}\equiv\hat{\mathcal{S}}^{(0)}={\sqrt{4\pi c_0 n\epsilon_0}}\int^\infty_0\!\!\! \mathrm{d}\omega \frac{\alpha^*_\mathrm{p}(\omega) \hat{a}_{s00}(\omega) +\text{H.c.}}{\sqrt{\hbar \omega}}\, .
\label{baseSN_1ch}
\end{equation}

For a non-displaced quantum state of the field, like the vacuum, the expectation values of signal operators  vanish. When squared, however, the signals lead to non-vanishing expectation values associated with their variance. Through unfolding of the Bogoliubov transformations, it is possible to see that, depending on $j$, the EO signals are functionals of either $\hat{a}_{s00}(\Omega)$ or $\hat{a}_{s00}(\omega)$. Apart from $\langle[\hat{\mathcal{S}}^{(j)}]^2\rangle$, this leads to crossterm contributions to the variance in the form $\langle\hat{\mathcal{S}}^{(j)}\hat{\mathcal{S}}^{(j+ 2)}\rangle$, $\langle\hat{\mathcal{S}}^{(j)}\hat{\mathcal{S}}^{(j+ 4)}\rangle$, and so on, as well as their conjugates. If one decomposes the total signal into terms depending solely on $\hat{a}(\Omega)$ and $\hat{a}(\omega)$, $\sum_{ \text{odd}\, j}\hat{\mathcal{S}}^{(j)}$ and $\sum_{ \text{even}\, j}\hat{\mathcal{S}}^{(j)}$, these two contributions would be effectively related via a two-mode squeezing involving one (nonmonochromatic) mode from each frequency range, MIR and NIR~\cite{Sho}. For this reason, the noise registered in the NIR is larger than the base level determined by $\langle \hat{\mathcal{S}}^2_{\mathrm{sn}}\rangle=N$.

The main contribution of the electric-field fluctuations in the MIR vacuum to the EO signal variance is  $\langle(\hat{\mathcal{S}}^{(1)})^{2}\rangle\propto N^2\int_{0}^\infty\!\mathrm{d}\Omega\; \Omega\,(n/n_{_\Omega}) |R(\Omega)|^2$, as shown in~\cite{Suppl_Mat} and experimentally endorsed by Ref.~\cite{vacuum_samp} (see also~\cite{Moskalenko}). The possibility that BA might play a role in this measurement, however, must also be considered, as pointed out in Ref.~\cite{benea}. The respective analysis is presented below.

Complementary to $\langle(\hat{\mathcal{S}}^{(1)})^{2}\rangle$, there is the crossterm between $\hat{\mathcal{S}}^{(2)}[\hat{E}^{(3)}_s(\omega)]$ and $\hat{\mathcal{S}}_\mathrm{sn}[\delta\hat{E}_s (\omega)]$, which does not contribute to the variance within the range of validity of our approximations. This term has not been accounted for in Refs.~\cite{Moskalenko, benea, Buhmann2020, Buhmann2021}, but its contribution can be neglected in the considered measurement regime. 
Interestingly, both $\langle\hat{\mathcal{S}}^{(2)}\hat{\mathcal{S}}^{(0)}+\hat{\mathcal{S}}^{(0)}\hat{\mathcal{S}}^{(2)}\rangle$ and $\langle \hat{\mathcal{S}}^{(1)}\hat{\mathcal{S}}^{(1)}\rangle$ share contributions from the 1st- and 2nd-order perturbations in the initial sampled state from a state-evolution perspective~\cite{Suppl_Mat}, some of which retain the MIR sector of the vacuum unchanged, while others correspond to populated MIR modes [cf. Fig.~\ref{fig1}(c)]. These perturbations are the source of EO signals and are comprised of a continuum of states in superposition with the initial state. As long as their $N$-dependent superposition coefficients are much smaller than the coefficient of the initial state, the 1st- and 2nd-order perturbations of the initial state are the dominating BA, whereas the measurement of the vacuum noise can be seen as effectively BA free. As the number $N$ of photons per probe pulse increases, further BA-induced contributions with coefficients growing even faster with $N$, such as the ones contributing to $\langle[\hat{\mathcal{S}}^{(2)}]^2\rangle$ and terms involving $\hat{\mathcal{S}}^{(j>2)}$, become significant in the quantum superposition of states, with our perturbation approach breaking down as they start to dominate.

To evaluate the BA effect on the measurement results, we derive all contributions from Eqs.~\eqref{E_NIR} and \eqref{E_MIR} to the EO signal variance up to 4th order. Once again, the crossterm between the SN signal and $\hat{\mathcal{S}}^{(4)}[\hat{E}^{(5)}(\omega)]$ vanishes within our approximations. The remaining $\hat{a}(\omega)$-dependent contribution, which also enhances the base SN, results from $\langle [\hat{\mathcal{S}}^{(2)}]^2\rangle$ and scales as $N^3$. Concurrent to this contribution is the $\hat{a}(\Omega)$-dependent (cross)term $\langle \hat{\mathcal{S}}^{(1)}\hat{\mathcal{S}}^{(3)}+\hat{\mathcal{S}}^{(3)}\hat{\mathcal{S}}^{(1)}\rangle$, which scales as $N^3$ but has an opposite sign to $\langle [\hat{\mathcal{S}}^{(2)}]^2\rangle$, leading therefore to a variance reduction. 

\begin{figure}
    \centering
    \includegraphics[width=\linewidth]{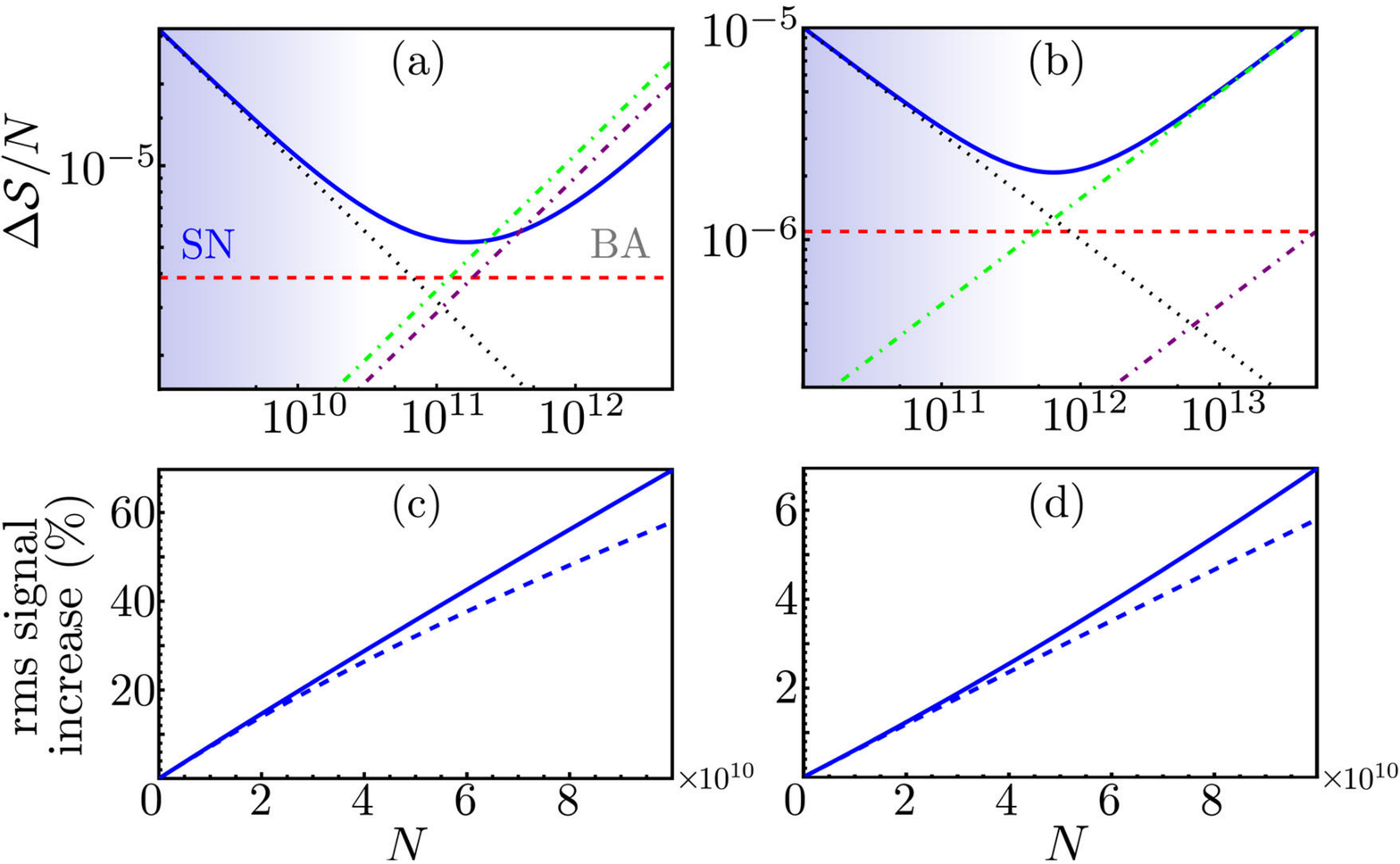}
 \caption{ (a), (b) Ratio $\Delta \mathcal{S}/N$ (with the rms signal $\Delta \mathcal{S}={\langle\hat{\mathcal{S}}^2\rangle}^{1/2}$) in dependence of the number of photons per probe pulse $N$. The solid blue lines represent the total rms signal per photon, the dotted black lines represent the base-SN contribution, the dashed red lines show the main EO rms signal and the green [purple] dot-dashed lines account for $\langle [\hat{\mathcal{S}}^{(2)}_{\text{eo}}]^2\rangle$ [$\langle \hat{\mathcal{S}}^{(1)}_{\text{eo}}\hat{\mathcal{S}}^{(3)}_{\text{eo}}+\hat{\mathcal{S}}^{(3)}_{\text{eo}}\hat{\mathcal{S}}^{(1)}_{\text{eo}}\rangle$]. The background gradient illustrates the transition between the effectively BA-free, shot-noise dominated (SN) and the BA-dominated (BA) regimes. (c), (d) Increase of $(\Delta \mathcal{S}-\Delta \mathcal{S}_{\mathrm{sn}})/\Delta \mathcal{S}_{\mathrm{sn}}$ with $N$. The solid (dotted) blue lines contain up to 4th(2nd)-order contributions. Left (right) plots correspond to parameter set 1 (set 2), see main text. 
  }\label{Noise}
\end{figure}

Figure~\ref{Noise} illustrates several aspects of the behavior of the EO signal as a function of $N$. For the sake of comparison, we provide plots corresponding to the experimental parameters of Refs.~\cite{Moskalenko} (set 1, left) and~\cite{benea} (set 2, right). For the probe pulses we assume $\omega^{(\text{set} 1)}_{\mathrm p}/(2\pi)=247$~THz and $\omega^{(\text{set} 2)}_{\mathrm p}/(2\pi)=375$~THz, and
spectral bandwidths of $\Delta\omega^{(\text{set} 1)}_{\mathrm p}/(2\pi)=150$~THz and $\Delta\omega^{(\text{set} 2)}_{\mathrm p}/(2\pi)=2.77$~THz with rectangular spectral shape and
flat phase. We consider beam waist
radii of $w^{(\text{set} 1)}_0=3~\mu$m and $w^{(\text{set} 2)}_0=125~\mu$m. For the NX we use $L^{(\text{set} 1)}=7~\mu$m and $L^{(\text{set} 2)}=3~$mm, $r_{_{41}}=3.9$~pm/V, $n^{(\text{set} 1)}=2.76$, $n^{(\text{set} 2)}=2.85$,
$n^{(\text{set} 1)}_{\mathrm{g}}=2.9$, $n^{(\text{set} 2)}_{\mathrm{g}}=3.18$ and $n_{_\Omega}$ varying  slightly within the relevant THz frequency range~\cite{Suppl_Mat}. We do not include contributions from four-wave mixing since they only affect the probe~\cite{Suppl_Mat}.

In the case of the total root-mean-square (rms) signal per probe photon shown in Figs.~\ref{Noise}(a) and (b), we see a considerable deviation from the result determined solely by the SN and the main EO contribution~\cite{Moskalenko} when photon numbers are larger than $\sim 10^{11}$, with minima at $N=1.6\times 10^{11}$ for set 1 [Fig.~\ref{Noise}(a)] and $N=6.4\times 10^{11}$ for set 2 [Fig.~\ref{Noise}(b)], roughly when $\langle [\hat{\mathcal{S}}^{(1)}]^2\rangle\sim \langle [\hat{\mathcal{S}}_{\mathrm{sn}}]^2\rangle$. Contrary to naive expectations that $\langle(\hat{\mathcal{S}}^{(1)})^{2}\rangle$ would generally dominate the total rms signal for large $N$, our results show that increasing the intensity of the probe pulse beyond a certain value has a rather detrimental effect, since both the base SN and main EO signal variance are rapidly overtaken by the cascaded effects. A reliable sampling of quantum states therefore can be realized for $N$ considerably smaller than its value at the minima of the solid blue curves in Figs.~\ref{Noise}(a) and (b). Figures~\ref{Noise} (c) and (d) show the (normalized) detected rms signal on top of the base SN contribution. Departure from zero allows for a clear visualization of the EO contributions, with the onset of the 4th-order terms observed as divergence between the solid and the dashed blue lines in each figure.
The threshold for the regime dominated by BA is analyzed in~\cite{Suppl_Mat}.

For a two-channel setup [Fig.~\ref{autocorrelation}(a)], the probe beam undergoes a beam-splitting operation before any of the aforementioned processes [corresponding to Fig.~\ref{fig1}(b)] take place. The two probe pulses released from a 50:50 beam splitter, although having the same intensity profiles (half of the input intensity each), carry not only different phases (reflected and transmitted beams differ in phase by $ \pi/2)$, but also commuting annihilation and creation operators due to the admixture of vacuum noise~\cite{Suppl_Mat}. Once the first probe pulse meets the NX, its interactions with the MIR vacuum will generate the BA contributions as discussed above. The second probe pulse will reach the NX with a time delay $\tau$ and interact not only with that MIR vacuum, but also with the BA contributions generated by the passage of the first probe, as well as generate its own BA contributions that can interact with the first probe. 
Each output field undergoes its own ellipsometry detection, and the respective signals from the two channels, $\hat{\mathcal{S}}_{\mathrm{ch1}}$ and $\hat{\mathcal{S}}_{\mathrm{ch2}}$, are then multiplied before readout, rendering a delay-dependent signal variance with the properties of a correlation function: $G(\tau)=\frac{1}{2C}\langle0|\{\hat{\mathcal{S}}_\mathrm{ch2}(\tau),\hat{\mathcal{S}}_\mathrm{ch1}(\tau)\}|0\rangle = \frac{1}{C}g(\tau)$ with $C=(n^3 L \omega_\mathrm{p} r_{41}N/c_0)^2$~\cite{benea}. Here and in what follows, the subscripts $\mathrm{ch1}$ and $\mathrm{ch2}$ shall describe quantities related to the channels 1 and 2, respectively. 

We consider a setup in which the directions of the central wave vectors of the beams in the two channels deviate only slightly from each other (i.e., $\vec{k}_\mathrm{ch1}\cdot \vec{k}_\mathrm{ch2}\approx k_\mathrm{ch1} k_\mathrm{ch2}$). This allows for consideration of effectively coplanar beam waists in the NX, as well as nearly collinear phase matching for the wave-mixing processes, therefore ensuring that the treatment of fields in terms of the paraxial decomposition is still well justified. In order to avoid considerable deviations of output wave vectors from either $\vec{k}_\mathrm{ch1}$ or $\vec{k}_\mathrm{ch2}$, only the second set of parameters will be considered. Due to the limited beam waist, mixing between fields from different channels during the ellipsometry step is avoided.  
For the quantum fluctuations of the probe pulses $[\hat{a}_\mathrm{ch1}(\omega),\hat{a}^\dagger_\mathrm{ch2}(\omega')]=0$, leading to
$\langle \{\hat{\mathcal{S}}^{(0)}_\mathrm{ch1},\hat{\mathcal{S}}^{(0)}_\mathrm{ch2}\}\rangle =0$ [cf.~\eqref{baseSN_1ch}], thus the two-channel equivalent of the base SN does not contribute to $g(\tau)$. In general, the (time-dependent) signal operators are given by equations similar to Eq.~\eqref{S_eo}, in which the cascaded contributions \eqref{E_NIR} and \eqref{E_MIR} (for $m\geq 2$) are now composed of convolutions with either $E_{\mathrm{p}, \mathrm{ch2}}(\omega)e^{-i(\omega \tau+\pi/2)}$ or $E_{\mathrm{p},\mathrm{ch1}}(\omega)$, splitting each single-channel contribution $\hat{E}^{(m)}_s$ into $2^{m-1}$ terms. Ellipsometry conducted with either of these fields then leads to $\hat{\mathcal{S}}^{(j)}_\mathrm{ch2}$ or $\hat{\mathcal{S}}^{(j)}_\mathrm{ch1}$, respectively (cf.~\cite{Suppl_Mat}). Since oscillations in $\tau$ with NIR frequencies can neither be resolved nor are of major interest in such an experiment, only contributions to $G(\tau)$ oscillating at MIR frequencies will be considered.
   
Fig.~\ref{autocorrelation}(b) shows $g(\tau)$ with all non-negligible terms up to 4th order. The main contribution, which depends on $\hat{E}^{(2)}_\mathrm{ch1}(\tau)\hat{E}^{(2)}_\mathrm{ch2}(\tau)$, is proportional to $ N^2\int_{0}^\infty\!\mathrm{d}\Omega\; \Omega\,(n/n_{_\Omega}) |R(\Omega)|^2\cos (\Omega\tau)$. This behavior is also seen in the measured data in Ref.~\cite{benea} up to differences related to the spectral shape chosen for the probe. Contributions from higher-order terms are minor up to $N\sim 10^{11}$. At such probe intensities, these terms have then the same order of magnitude as the 2nd-order terms. For $N\sim 10^8$, as utilized in the experiments~\cite{subcycle, benea, benea_vac},  higher-order contributions to $G(\tau)$ are negligible.
 
 The characteristic time scale of $G(\tau)$ has a duration similar to that of the probe pulses, since only when the two probes share interactions with the same propagating MIR modes (i.e., when there is some overlap between their interaction time windows) the multiplied signal will not vanish
 . The oscillations of $G(\tau)$ in Fig.~\ref{autocorrelation}(b) happen with a time scale being approximately the inverse of the average probed MIR (angular) frequency and reflect the interference between modes from different channels
 . From a state-evolution perspective, $G(\tau)$ includes interchannel probe-probe correlations mediated by MIR states, populated or not~\cite{Suppl_Mat}.

\begin{figure}[t!]
    \centering
    \includegraphics[width=0.88\linewidth]{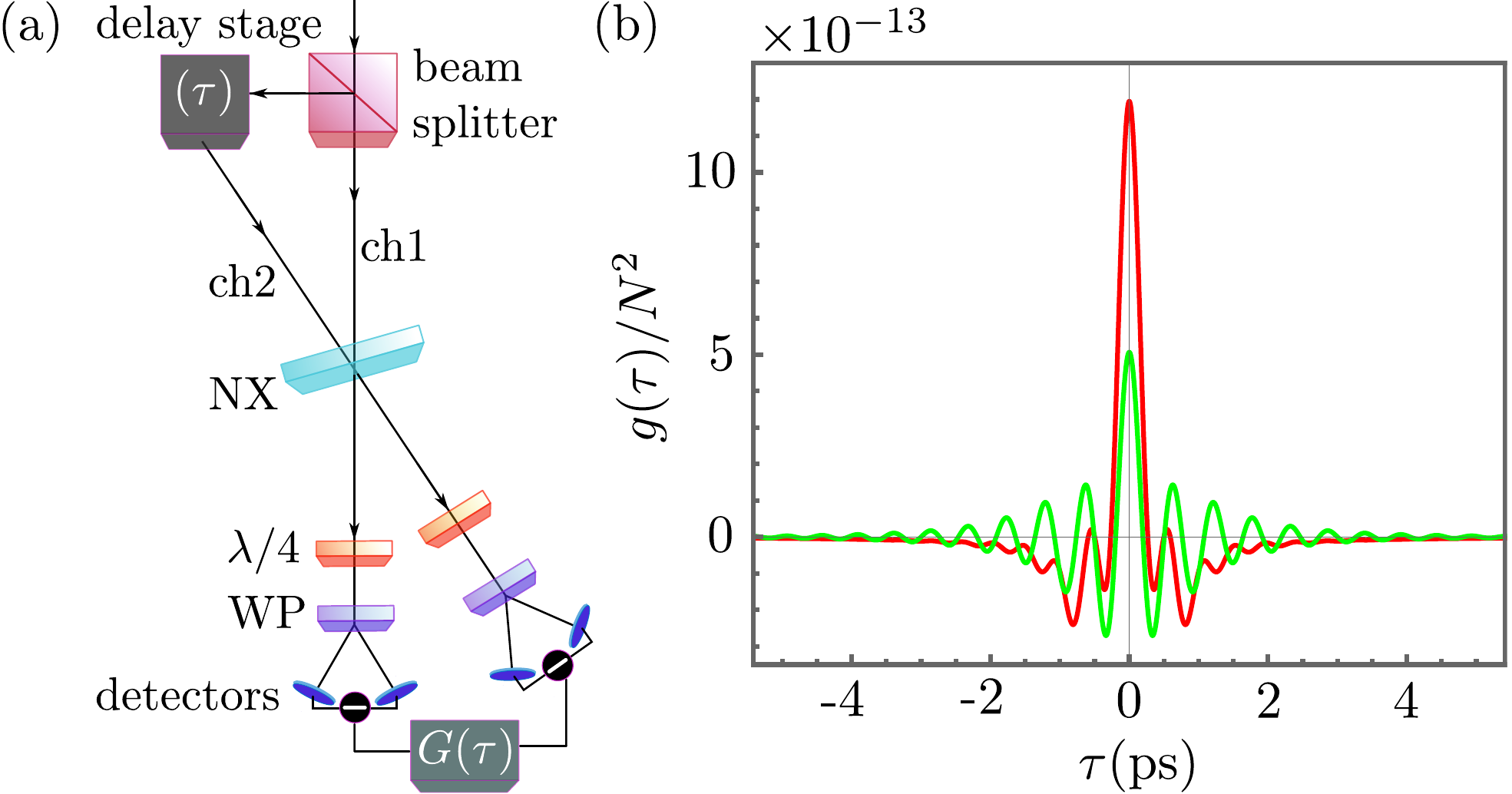}
 \caption{(a) Illustration of an EO measurement with two channels ($\mathrm{ch1}$ and $\mathrm{ch2}$), with $\lambda/4$, WP and $G(\tau)$ representing the quarter-wave plate, the Wollaston prism and the correlation register. (b) Total 2nd- (red) and 4th-order (green) contributions to $g(\tau)$ for $N=10^{11}$, representing signal autocorrelations in the two-channel measurements.}
  \label{autocorrelation}
\end{figure}

 In summary, our results show that in single-channel EO measurements of the electromagnetic vacuum there is a suitable setup-dependent range of probe-field intensities to minimize the contributions from generated MIR photons to the signal. For small (large) probe intensities, the results are inevitably contaminated by SN (BA). The BA to the MIR states is inherent to the EO measurement process but the measurement result for the detected MIR signal variance can be considered effectively BA-free (weak measurement) as long as the perturbations of the MIR states remain relatively small. Furthermore,as shown above, the base SN can be evaded by the use of two channels, and the BA starts to affect the measured correlations in the MIR vacuum for probe intensities several orders of magnitude above the experimentally used values. Some further BA evasion might be achieved by usage of a proper post-selection scheme after the NX, and future works might explore the possibilities offered by this approach.

\vspace{0.2cm}
\begin{acknowledgments}

T.L.M.G., A.L., A.S.M. and G.B. acknowledge
funding by the Deutsche Forschungsgemeinschaft
(DFG) - Project No. 425217212 - SFB 1432.
T.L.M.G. and A.S.M. gratefully acknowledge
the funding by the Baden-W\"{u}rttemberg Stiftung via
the Elite Programme for Postdocs. A.S.M. was also
supported by the National Research Foundation of Korea
(NRF) grant funded by the Korean government (MSIT)
(2020R1A2C1008500).  I.V. acknowledges support by the Institute for Basic Science (Project number IBS-R024-D1). We thank P. Sulzer for the productive discussions at the early stage of this research.
\end{acknowledgments}

\onecolumngrid
\clearpage

\setcounter{equation}{0}
\setcounter{figure}{0}
\setcounter{table}{0}
\setcounter{page}{1}

\makeatletter

\renewcommand{\theequation}{S\arabic{equation}}
\renewcommand{\thefigure}{S\arabic{figure}}
\renewcommand{\bibnumfmt}[1]{[S#1]}
\renewcommand{\citenumfont}[1]{S#1}

\setcounter{secnumdepth}{3}
\renewcommand\thesection{\arabic{section}}

\begin{center}
\textbf{\large \underline{Supplemental Material}}\\[0.5cm]
\textbf{\large Back action in quantum electro-optic sampling of electromagnetic vacuum fluctuations}
\end{center}

\begin{center}
T. L. M. Guedes, I. Vakulchyk, D. V. Seletskiy, A. Leitenstorfer, A. S. Moskalenko, and G. Burkard
\end{center}

\section{Higher-order contributions to the signal variance}\label{sup_sec_1}

In this section, we provide the expressions for the higher-order contributions (in terms of powers of the susceptibility or of the probe photon number) to the signal variance. All equations have been derived with the aid of Eqs.~\eqref{E_NIR}-\eqref{S_eo} and intermediary steps will not be provided in order to avoid lengthy derivations. 

Apart from the main contribution to the signal variance given by
\begin{equation}
  \langle(\hat{\mathcal{S}}^{(1)})^{2}\rangle\!=\!N^2\left(\!n^3\frac{L \omega_\mathrm{p}}{c_0}r_{_{\!41}}\!\!\right)^{\!\!2}\,
  \frac{\hslash \int_{0}^\infty\!\mathrm{d}\Omega\; \Omega\,
(n/n_{_\Omega}) |R(\Omega)|^2}{4\pi^2\epsilon_0c_0n w_0^2},
\label{signal1}
\end{equation}
there is another contribution of second order in $r_{41}$ arising from the crossterm between the SN signal and the signal arising from the measurement of the field $\hat{E}^{(3)}_s(\omega)$ (i.e., the field generated when the nonlinear mixing with the probe takes the NIR vacuum field to the MIR and this new generated MIR field mixes then once again with the probe to give a NIR field). The respective contribution to the signal variance is given by
\begin{equation}
  \langle\{\hat{\mathcal{S}}^{(2)},\hat{\mathcal{S}}^{}_\mathrm{sn}\}\rangle\!=\! -\frac{2C_1}{\sqrt{3}}\,
 \int_{-\infty}^\infty\!\mathrm{d}\Omega\; \Omega\,
                                         (n/n_{_\Omega}) \big[i\zeta_{\omega,\Omega}f_-(\Omega)R^*(\Omega)/(d\frac{L \omega}{2c_0 n })\big],
\label{signalcross_ch1} 
\end{equation}
where 
 \begin{equation}\label{Eq:C_i}
   C_j=N^{j+1}\left(\!n^3\frac{L \omega_\mathrm{p}}{c_0}r_{_{\!41}}\!\!\right)^{2j} \left(\frac{ \hbar }{4\pi^2 \epsilon_0 c_0 n w^2_0}\right)^j
 \end{equation}
 and $j\in \mathbb{N}$.
The spatial integral for $\hat{\mathcal{S}}^{(2)}$ (with $w'_0=w_0/\sqrt{3}$) gives $A^{(3)}=(2/\sqrt{3\pi^2})/w^2_0$. Note that the variable $\omega^{-1}$ in Eq.~\eqref{signalcross_ch1} gets canceled by the $\omega$ in $\zeta_{\omega, \Omega}$, so that the result does not depend on it. For our specific choice of the probe profile, $f_-(\Omega)=F(\Omega)$ within the range of MIR frequencies selected by the phase matching, making the integrand an even function of $\Omega$. Hence, the right-hand side of Eq.~\eqref{signalcross_ch1} vanishes.

Before delving into the 4th-order contributions to the signal variance, it is worth introducing some building-block functions that will be used through the remainder of the Supplemental Material:
\begin{equation}
R^{(\pm)}_i(\Omega, X,\tau)=\frac{1}{2} \mathrm{sinc}\!\left[\frac{L\Omega}{2c_0}(n_{_\Omega}-n_\mathrm{g})\right]e^{\frac{iL\Omega}{2c_0}(n_{_\Omega}-n_\mathrm{g})}\frac{\int^\infty_{-\infty}\mathrm{d}\omega\, \alpha^*_\mathrm{p} (\omega)\alpha_\mathrm{p}(\omega-\Omega)K^{(\pm)}_i(\omega, X,\tau)}{\int^\infty_{0}\!\!\!\mathrm{d}\omega\,|\alpha_\mathrm{p} (\omega)|^2 } \, ,
\label{R}
\end{equation}

\begin{equation}
W^{(\pm)}_i(\Omega, \Omega', X, \tau)=\frac{\int^\infty_{-\infty} \!\!\!\mathrm{d}\omega\,\omega \alpha^*_\mathrm{p} ( \omega-\Omega) K^{(\pm)}_i(\omega, X,\tau )\alpha_\mathrm{p} (\omega-\Omega')}{\int^\infty_{0}\!\!\!\mathrm{d}\omega\,|\alpha_\mathrm{p} (\omega)|^2 } \, ,
\label{W}
\end{equation}

\begin{equation}
\begin{aligned}
G^{(\pm, \pm')}_{ij}( \Omega, \Omega', X, Y , \tau)=\left(d\frac{L\omega^{3/2}_\mathrm{p}}{2c_0 n}\right)^{-2}\!\!\!\!\zeta^*_{\omega_\mathrm{p}, \Omega} \zeta_{\omega_\mathrm{p}, \Omega'}
 R^{(\pm)}_i (\Omega, X, \tau)W^{(\pm')}_j(\Omega, \Omega', Y, \tau) \, .
\end{aligned}
\end{equation}
The functions $K^{(\pm)}_i$ can have different shapes depending on the choice of the index $i$, namely $K^{(\pm)}_0 (\omega,X,\tau) =\theta (\pm \omega)$, $K^{(\pm)}_1(\omega, X,\tau)=\cos[\tau(\omega+ X)]\theta (\pm \omega)$ and $K^{(\pm)}_2(\omega, X,\tau)=\sin[\tau(\omega+ X)]\theta(\pm\omega )$. Here $\theta(x)$ denotes the
Heaviside step function. In the trivial $i=0$ case, we have
\begin{equation}
\sum_{s=\pm} R^{(\pm)}_0(\Omega)=\mathrm{sinc}\!\left[\frac{L\Omega}{2c_0}(n_{_\Omega}-n_\mathrm{g})\right]\mathrm{exp}\left[\frac{iL\Omega}{2c_0}(n_{_\Omega}-n_\mathrm{g})\right](1-|\Omega|/\Delta \omega)\theta (1-|\Omega|/\Delta \omega) = R(\Omega) \, .
\end{equation}

There are three 4th-order contributions to the signal variance. One of them is just the square of the signal associated with  $\hat{E}^{(3)}_s(\omega)$ field mentioned above (i.e., shot-noise enhancement). The corresponding expression is given by
\begin{equation}
  \langle(\hat{\mathcal{S}}^{(2)})^{2}\rangle\!=\! 
  \frac{C_2}{3}
\sum_{t=\pm}\int^\infty_{-\infty} \!\!\!\!\!\mathrm{d}\Omega \!\int^\infty_{-\infty}\!\!\!\!\!  \mathrm{d}\Omega' (\Omega\Omega') R^{*} (\Omega')G^{(t,+)}_{00} (\Omega, \Omega',0,0, 0)\, .
\label{signalNIRxNIR}
\end{equation}
The second term of same perturbative order results from the mixing between the main signal, $\hat{\mathcal{S}}^{(1)}[\delta \hat{E}_s(\Omega)]$, and the signal $\hat{\mathcal{S}}^{(3)}[ \hat{E}^{(3)}_s(\Omega)]$, giving 
\begin{equation}
\langle \{\hat{\mathcal{S}}^{(3)},\hat{\mathcal{S}}^{(1)}\}\rangle = \frac{-C_2}{\sqrt{2}}
\sum_{t,s=\pm}\int^\infty_{-\infty} \!\!\!\!\!\mathrm{d}\Omega \!\int^\infty_0\!\!\!\!\!  \mathrm{d}\Omega' (\Omega\Omega') R^{*} (\Omega')G^{(t,s)}_{00} (\Omega, \Omega',0,0, 0)\frac{n}{n_{\Omega'}}.
\label{signalx}
\end{equation}
Here, $A^{(4)}=\sqrt{2/\pi^3}/w^3_0$ for $\hat{\mathcal{S}}^{(3)}$ with $w'_0=w_0/2$.
The last contribution comes from the crossterm between the SN and the signal $\hat{\mathcal{S}}^{(4)}[ \hat{E}^{(5)}_s(\omega)]$ (with $A^{(5)}=4/(\sqrt{5}\pi^2w^2_0)$ for $w'=w_0/\sqrt{5}$). This term gives zero variance contribution for our choice of the probe profile, similarly to Eq.~\eqref{signalcross_ch1} [because $\sum_t W^{(t)}_0(\Omega,\Omega')=-\sum_t W^{(t)}_0(-\Omega,-\Omega')$]:
\begin{equation}
  \langle\{\hat{\mathcal{S}}^{(4)}_\mathrm{eo},\hat{\mathcal{S}}^{}_\mathrm{sn}\}\rangle\!=
  \frac{C_2}{\sqrt{5}}
\sum_{t,s=\pm}\int^\infty_{-\infty} \!\!\!\!\!\mathrm{d}\Omega \!\int^\infty_{-\infty}\!\!\!\!\!  \mathrm{d}\Omega' (\Omega\Omega') R^{(+)*}_{0} (\Omega', 0, 0)G^{(t,s)}_{00} (\Omega, \Omega',0,0, 0) \, .
\label{signalcross4}
\end{equation}

\section{Influence of the beam waist}

\begin{figure}[h]
    \centering
    \includegraphics[width=\linewidth]{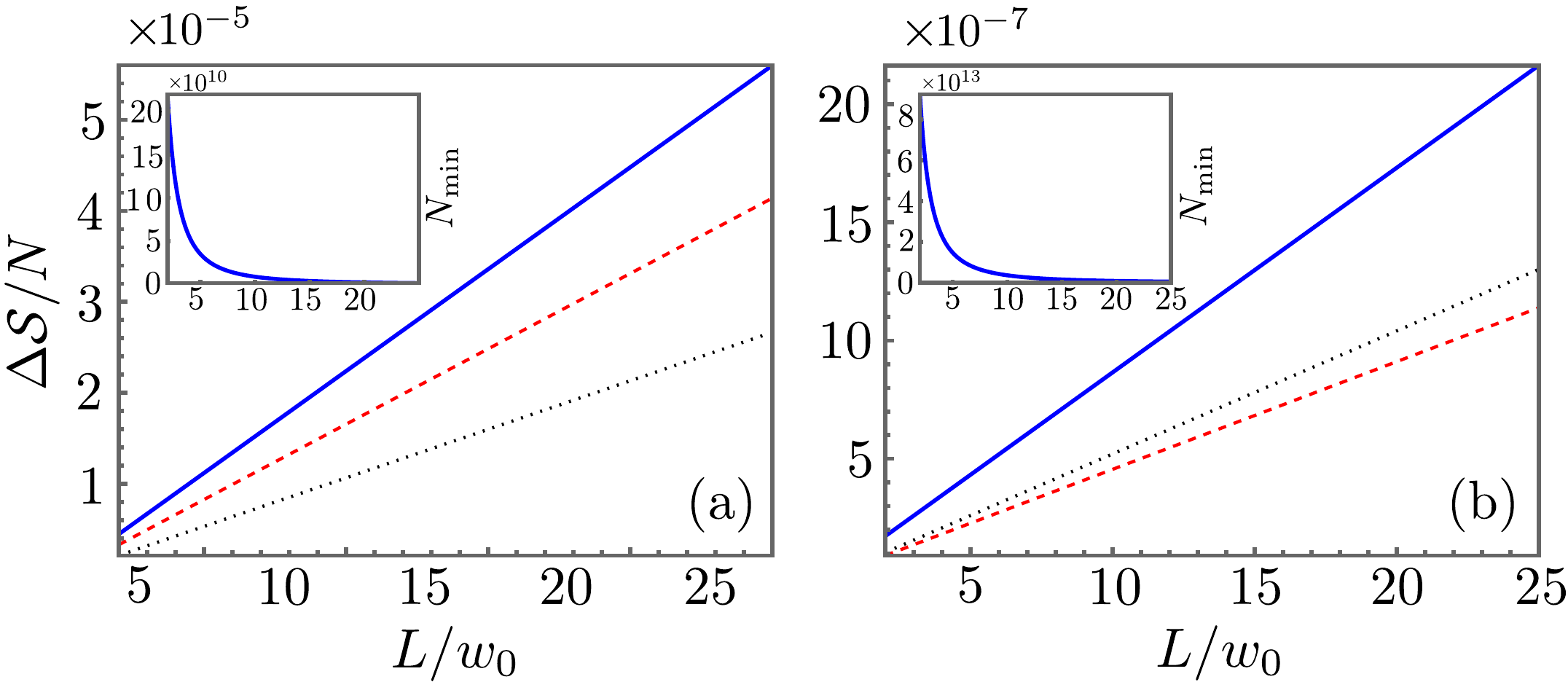}
  \caption{(a), (b) Minimum of $\Delta \mathcal{S} /N$ as a function of $w_0$. The solid blue line shows the total $\Delta {\mathcal{S}} /N$ value. The dotted black (dashed red) lines represent the SN (main EO) contributions. The insets show how the probe photon number that minimizes $\Delta \mathcal{S} /N$, $N_\text{min}$, varies as a function of $w_0$ (also normalized as $L/w_0)$. Left (right) plots correspond to parameter set 1 (set 2), apart from the value of $w_0$. 
}\label{photonnumber}
\end{figure}

The threshold for the regime dominated by BA is illustrated in Fig.~\ref{photonnumber} in dependence of the beam waist radius $w_0$. In Figs.~\ref{photonnumber}(a) and (b), one can see how the minimum of the total rms signal per photon (at the minimizing $N$ value, $N_\mathrm{min}$) depends on $w_0$ [normalized by $L^{(\text{set}1)}$ and $L^{(\text{set}2)}$, respectively]. For comparison, we also show the dependences of the SN and main ($j=1$) rms signals per photon on $w_0$ at $N=N_\mathrm{min}$. One can see that $\Delta {\mathcal{S}} /N (N_\mathrm{min}, L/w_0)$ increases linearly with $L/w_0$ and its slope is larger than the ones for both the SN and the main EO contributions, which also grow linearly. One can therefore expect that increasing the length-to-waist ratio makes the detection of the vacuum signal more difficult.

\section{Product signal with two probe beams}

In this section we discuss in more details the subtleties of the theoretical description in case of the measurement scheme implemented in Ref.~\cite{beneaSup}. 
One of the key features of this experiment is the use of a beam splitter to convert a single probe pulse into two pulses of equal durations. The beam splitter mixes the coherent probe of classical amplitude $\alpha_\mathrm{p}$ and its quantum vacuum contribution (described by operators $\hat{a}$) with the vacuum noise (represented by operators $\hat{a}_\text{free}$) accessing the classically free port of the device. The resulting transformation of the incoming probe field is given by
\begin{equation}
\begin{aligned}
&\hat{a}^{(\text{out})}_{1}(\omega) =T(\omega)\hat{a}^{(\text{in})}_{1}(\omega)+R'(\omega)\hat{a}^{(\text{in})}_{2}(\omega)= T(\omega)\alpha_\mathrm{p} (\omega)+ \big[T(\omega)\hat{a}(\omega)+R'(\omega)\hat{a}_{\text{free}}(\omega)\big]=T(\omega)\alpha_\mathrm{p} (\omega)+\hat{a}_\mathrm{p,ch1}(\omega)\, , \\
&\hat{a}^{(\text{out})}_{2}(\omega) =R(\omega)\hat{a}^{(\text{in})}_{1}(\omega)+T'(\omega)\hat{a}^{(\text{in})}_{2}(\omega)= R(\omega)\alpha_\mathrm{p} (\omega)+ \big[R(\omega)\hat{a}(\omega)+T'(\omega)\hat{a}_{\text{free}}(\omega)\big]=R(\omega)\alpha_\mathrm{p} (\omega)+\hat{a}_\mathrm{p,ch2}(\omega) \, .
\end{aligned}
\label{beamsplitter}
\end{equation}
It is worth mentioning that this transformation applies for all frequencies, including the MIR frequency range, in which $\alpha_\mathrm{p}(\omega)=0$. The new annihilation operators, $\hat{a}_\mathrm{p, ch1}$ and $\hat{a}_\mathrm{p, ch2}$, are associated with the two channels employed in the experiment. One interesting feature of the output operators is their commutativity: $[\hat{a}_\mathrm{p, ch1},\hat{a}^\dagger_\mathrm{p, ch2}]=TR^*+R'T'^*=0$ under the assumption of a dissipationless beam splitter. As a result, the product of two signals, each of which is a functional of beam-splitter creation and annihilation
output operators originating from a different channel, has vanishing expectation value. A less pronounced but also important feature is the phase introduced by the beam splitter on the reflected components of the fields. For the sake of agreement with the particular experimental realization, we shall consider an ideal 50:50 beam splitter without dispersion, with $T=T'=1/\sqrt{2}$ and $R=R'=i/\sqrt{2}$.

To account for the relative delay between modes in different channels, we consider the channel-two modes to be time shifted by a delay $\tau$ relative to the channel-one modes. This effectively introduces a factor $e^{-i\Lambda \tau}$ on the positive-frequency components of the channel-two fields when working in the frequency domain. Assuming that the fields from both channels meet at the NX in a quasi-parallel configuration (i.e., the angle between them is $\lesssim 5^\circ$), so that cross-section and propagation-length mismatch can be neglected at the focus of the beams and the frequency decomposition of fields can substitute the wave-vector one, wave mixing will involve contributions from both channels. Hence the $\tau$-dependent phase (with a suitable frequency argument) can appear at any (if not several) steps when applying Eqs.~\eqref{E_NIR} and \eqref{E_MIR}. There are, however, constraints on the possibilities of combination of phase factors in the convolutions of field operators with (time-shifted) probe pulses, based on the actual spatial distribution of output wave vectors. Since $\vec{k}_\Omega$ effectively does not affect the direction of $\vec{k}_{\omega, \mathrm{ch}1/2}$, so that $\vec{k}_{\omega, \mathrm{ch1/2}}\pm\vec{k}_\Omega\approx\vec{k}_{\omega\pm \Omega,\mathrm{ch}1/2}$ for either channel, combinations of fields from different channels should be such that the sum of wave vectors involved in the nested convolutions aligns with the input wave vector of one of the channels: $\sum_i \vec{k}_i\approx\vec{k}_\mathrm{ch1/2}$. This leads to a set of effective selection rules:  
\begin{itemize}
\item convolution of two field contributions from the same channel [e.g., by consecutive application of Eqs.~\eqref{E_NIR} and \eqref{E_MIR} with the same probe] does not contribute to the final wave vector, since such a same-channel convolution involves two wave vectors that are opposite to each other;
\item for fields that are functionals of the NIR annihilation and creation operators, contributions to the expectation value of a product of two signal operators will not vanish only when both multiplied signals depend on operators from the same channel --- the output in one of the channels should carry NIR annihilation and creation operators from the other channel or, equivalently, the sum of wave vectors corresponding to a given $\hat{E}^{(m)}_{s,\mathrm{ch2/1}}(\omega)$ should take the input vector $\vec{k}_\mathrm{ch1/2}$ of $\delta\hat{E}_{s,\mathrm{ch1/2}}(\omega)$ to the output $\vec{k}_\mathrm{ch2/1}$;
\item only odd total numbers of $\tau$-dependent phase factors distributed between the nested convolutions in both $\hat{E}^{(m)}_{s,\mathrm{ch1}}(\omega)$ and $\hat{E}^{(m')}_{s,\mathrm{ch2}}(\omega)$ lead to cross-signals with oscillations at MIR frequencies.
\end{itemize}
The ellipsometry step takes place independently for each channel, with channel-one fields being superposed with $E_\mathrm{p, ch1}$ and channel-two fields with $E_\mathrm{p, ch2}$, respectively. For this reason, the base SN signals in each channel commute with each other [these signals in channels one and two are proportional to integrals over $\hat{a}_\mathrm{p, ch1}(\omega)$ and $\hat{a}_\mathrm{p, ch2}(\omega)$, respectively, as well as their corresponding conjugates]. This commutation relation is what allows the measurement of the product of signals from different channels to have its base SN contribution reduced to negligible values over a large enough averaging sample, at least as long as SN enhancement does not become appreciable. 
Upon multiplication of the signals from the two channels, each signal-variance component treated in the previous section will now give rise to a plethora of time-dependent components, many of which oscillate with frequencies in the bandwidth of the probe. The latter components will not be considered in this work since with their high-frequency oscillations they effectively just average out to zero in the discussed experiment. With the purpose of comparison, we shall consider the beam-splitter output probes to have the same intensity as the probes used in the single-channel calculations.

\section{Autocorrelation functions}

Upon application of the selection rules introduced in the previous section it is possible to reduce the total of 56 contributions to the correlation function (up to 4th order) to just a few. At 2nd order, only 2 out of the 8 possible outcomes fulfil the selection rules. We start by the description of the mainly-contributing 2nd-order term, the squared MIR vacuum signal of the form:
\begin{equation}
  \frac{1}{2}\langle\hat{\mathcal{S}}^{(1)}_\mathrm{ch1}\hat{\mathcal{S}}^{(1)}_\mathrm{ch2}+\hat{\mathcal{S}}^{(1)}_\mathrm{ch2}\hat{\mathcal{S}}^{(1)}_\mathrm{ch1}\rangle (\tau)\!=\! C_1
  \int_{-\infty}^\infty\!\mathrm{d}\Omega\; \Omega\,
                                         (n/n_{_\Omega}) \big|R(\Omega)\big|^2\cos(\Omega \tau)\;.
\label{signalcross} 
\end{equation}
Its time dependence comes from the delay in one of the channels, introducing an $e^{-i(\omega-\Omega)\tau}$ phase factor on the (channel-two) field that mixes with the MIR vacuum in Eq.~\eqref{E_NIR} and an $e^{-i\omega\tau}$ on the probe that takes part in the (channel-two) ellipsometry, resulting in the overall $\Omega\tau$ dependence. The other contribution of 2nd order reads:
\begin{align}
  \frac{1}{2}\langle\hat{\mathcal{S}}^{(2)}_\mathrm{ch1}\hat{\mathcal{S}}^{(0)}_\mathrm{ch2}+\hat{\mathcal{S}}^{(0)}_\mathrm{ch1}\hat{\mathcal{S}}^{(2)}_\mathrm{ch2}+\hat{\mathcal{S}}^{(2)}_\mathrm{ch2}\hat{\mathcal{S}}^{(0)}_\mathrm{ch1}+\hat{\mathcal{S}}^{(0)}_\mathrm{ch2}\hat{\mathcal{S}}^{(2)}_\mathrm{ch1}\rangle(\tau)&\!=\! \frac{-2C_1}{\sqrt{3}}
   \int_{-\infty}^\infty\!\mathrm{d}\Omega\; \Omega\,
                                     (n/n_{_\Omega}) \big[i\zeta_{\omega,\Omega}f_-(\Omega)R^*(\Omega)\cos(\Omega \tau)/(d\frac{l \omega}{2c_0 n })\big]\;.
\label{signalcrossNIR} 
\end{align}
This lowest-order cross-talk contribution results when the NIR electric-field fluctuations in one channel are down-converted to the MIR and then up-converted to the NIR at another channel. Under our approximations, this contribution vanishes.
 
On top of the cross-signal contributions \eqref{signalcross} and \eqref{signalcrossNIR}, there are still contributions of higher perturbative orders. The next-order terms depend on $N^3$, and eventually overcome the $N^2$ terms as the probe-pulse intensity is increased. These $N^3$-dependent terms are similar in nature to the ones derived for the single-channel case, but differ in the many ways in which the probe pulses from the two channels can combine with the quantum components of the electromagnetic field. We shall therefore present them in the same order as we did in section~\ref{sup_sec_1}.

The 4th-order contribution related to the $E^{(3)}_s(\omega)$ field can be split into three parts: $\frac{1}{2} \langle \hat{\mathcal{S}}^{(2)}_\mathrm{ch1}\hat{\mathcal{S}}^{(2)}_\mathrm{ch2}+\hat{\mathcal{S}}^{(2)}_\mathrm{ch2}\hat{\mathcal{S}}^{(2)}_\mathrm{ch1}\rangle(\tau)\!=\! V^{(a)}_{\{2,2\}}(\tau) + V^{(b)}_{\{2,2\}}(\tau) +V^{(c)}_{\{2,2\}}(\tau)$. Here, $V^{(a)}_{\{2,2\}}$ results from two equivalent processes: one in which the output of wave vector $\vec{k}_\mathrm{ch2}-\vec{k}'_\mathrm{ch2}+\vec{k}''_\mathrm{ch2}$ is multiplied with the output corresponding to $\vec{k}_\mathrm{ch2}-\vec{k}'_\mathrm{ch2}+\vec{k}''_\mathrm{ch1}$ and another in which the $\mathrm{ch1}$ and $\mathrm{ch2}$ subscripts are swapped \footnote{In the schematic wave-vector description $\vec{k}_\mathrm{ch2}-\vec{k}'_\mathrm{ch2}+\vec{k}''_\mathrm{ch1}$, the unprimed vector $\vec{k}_\mathrm{ch2}$ represents the wave vector of the NIR operator $\delta\hat{E}_\mathrm{ch2}(\omega)$, while the primed and doubly primed vectors stand for the wave vectors of the probe pulses participating in the first and second convolutions, respectively, that generate the output field contribution.}.  $V^{(b)}_{\{2,2\}}$ originates from the product of field contributions with wave vectors $\vec{k}_\mathrm{ch1}-\vec{k}'_\mathrm{ch2}+\vec{k}''_\mathrm{ch2}$ and $\vec{k}_\mathrm{ch1}-\vec{k}'_\mathrm{ch1}+\vec{k}''_\mathrm{ch2}$, while  $V^{(c)}_{\{2,2\}}$ results from the product of contributions with wave vectors $\vec{k}_\mathrm{ch2}-\vec{k}'_\mathrm{ch2}+\vec{k}''_\mathrm{ch1}$ and $\vec{k}_\mathrm{ch2}-\vec{k}'_\mathrm{ch1}+\vec{k}''_\mathrm{ch1}$. Explicitly we find the following expressions:
\begin{equation}
V^{(a)}_{\{2,2\}}(\tau)= \frac{C_2}{3} \sum_{t,s=\pm}\int^\infty_{-\infty} \!\!\!\!\!\mathrm{d}\Omega \!\int^\infty_{-\infty}\!\!\!\!\!  \mathrm{d}\Omega' (\Omega\Omega') R^{(s)*}_{0} (\Omega', 0, 0)G^{(t,+)}_{00} (\Omega, \Omega',0,0, 0)[\cos(\Omega\tau)+\cos(\Omega'\tau)],
\label{V22a}
\end{equation}
 \begin{equation}
V^{(b)}_{\{2,2\}}(\tau)= \frac{C_2}{3} 
\sum_{s=\pm}\int^\infty_{-\infty} \!\!\!\!\!\mathrm{d}\Omega \!\int^\infty_{-\infty}\!\!\!\!\!  \mathrm{d}\Omega' (\Omega\Omega') [R^{(s)*}_{0} (\Omega', 0, 0)G^{(+,+)}_{11} (\Omega, \Omega',\Omega',0, \tau)+R^{(s)*}_{0} (\Omega', 0, 0)G^{(+,+)}_{22} (\Omega, \Omega',\Omega',0, \tau)],
\label{V22b}
\end{equation}
and
 \begin{equation}
V^{(c)}_{\{2,2\}}(\tau)= \frac{C_2}{3} 
\sum_{s=\pm}\int^\infty_{-\infty} \!\!\!\!\!\mathrm{d}\Omega \!\int^\infty_{-\infty}\!\!\!\!\!  \mathrm{d}\Omega' (\Omega\Omega') [R^{(+)*}_{1} (\Omega', \Omega, \tau)G^{(s,+)}_{01} (\Omega, \Omega',0,0, \tau)+R^{(+)*}_{2} (\Omega', \Omega, \tau)G^{(s,+)}_{02} (\Omega, \Omega',0,0, \tau)].
\end{equation} 
The behavior of each of the above contributions in dependence on the time delay $\tau$ is illustrated in Fig.~\ref{V22_independentplots}.

\begin{figure}[h!]
    \centering
    \subfigure{\includegraphics[width=7cm]{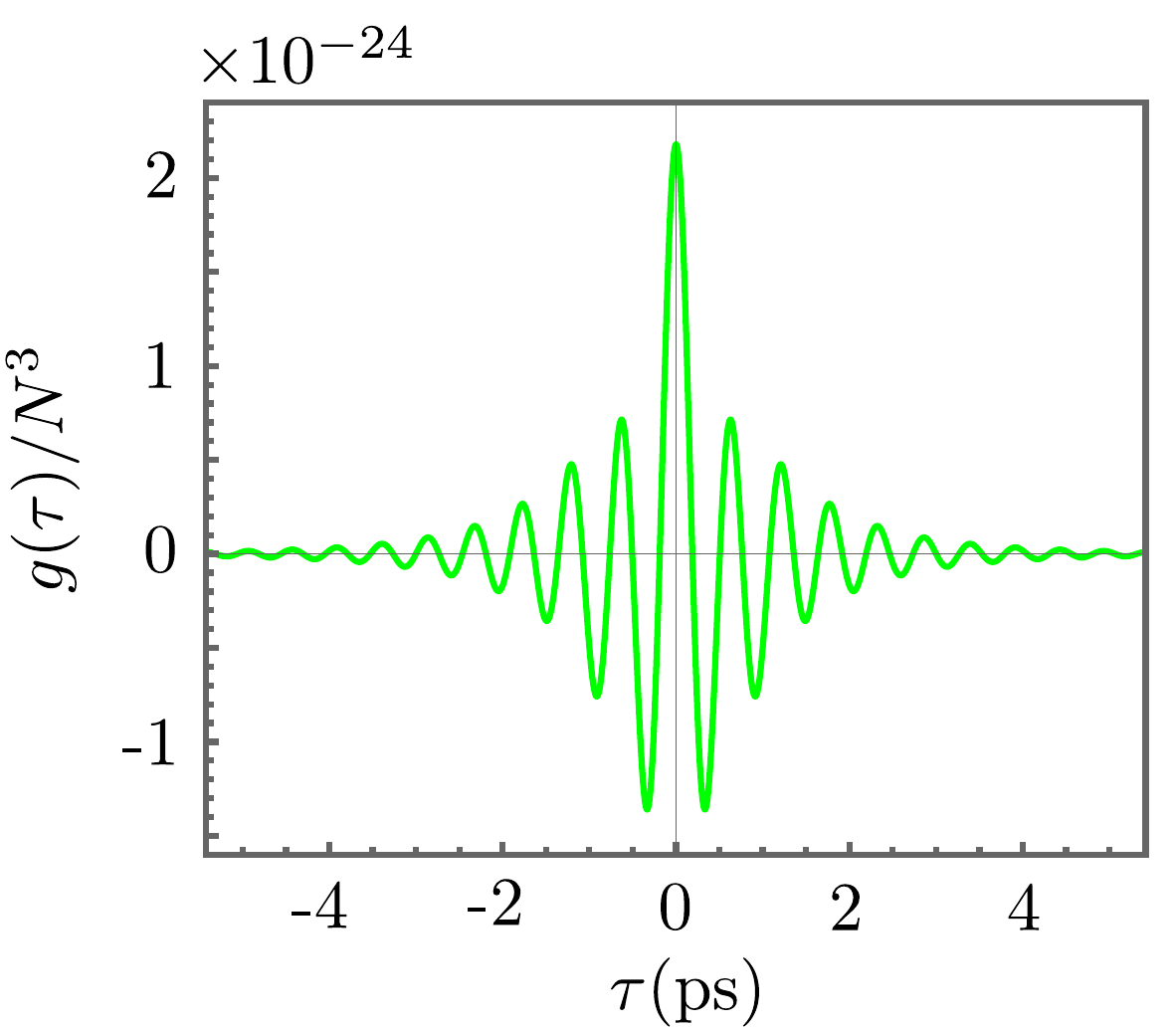}}\quad
\subfigure{\includegraphics[width=7cm]{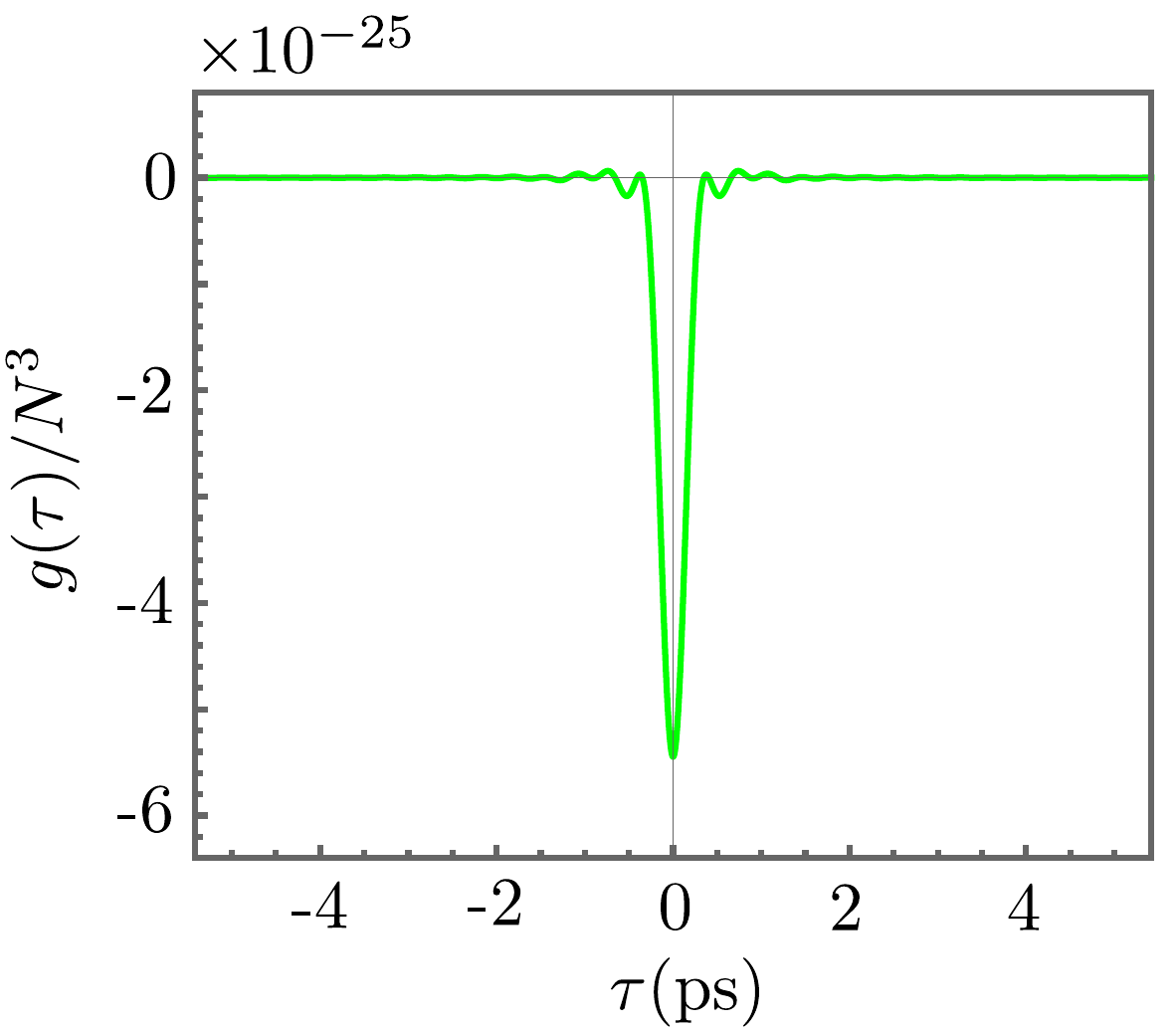}}
\end{figure} 
\begin{figure}[h!]
    \centering
    \includegraphics[width=7cm]{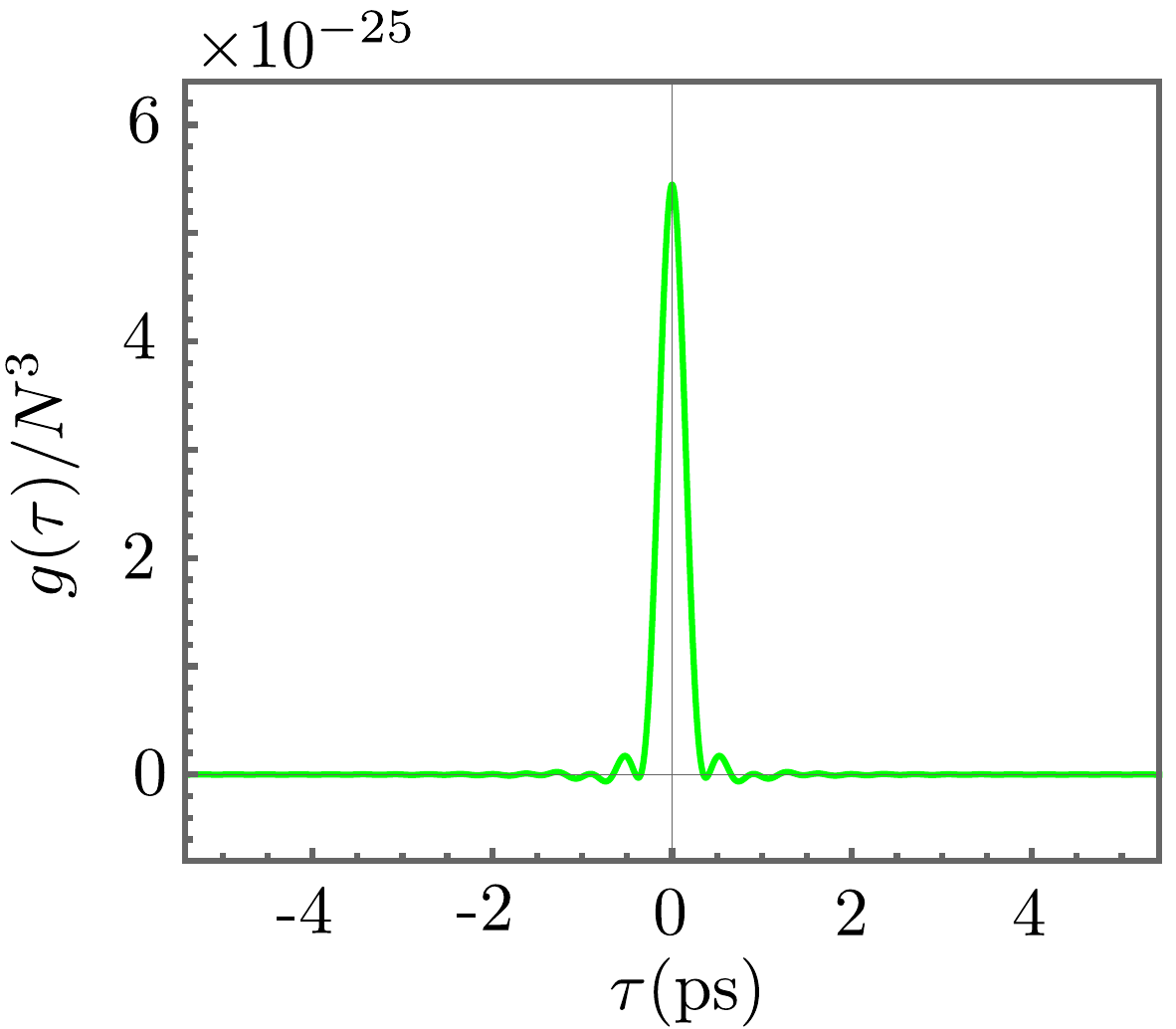}
 \caption{Top left: $V^{(a)}_{\{2,2\}}(\tau)/N^3$. Top right: $V^{(b)}_{\{2,2\}}(\tau)/N^3$. Bottom: $V^{(c)}_{\{2,2\}}(\tau)/N^3$.}
  \label{V22_independentplots}
\end{figure} 

\begin{figure}
\centering
\subfigure{
    \includegraphics[width=7cm]{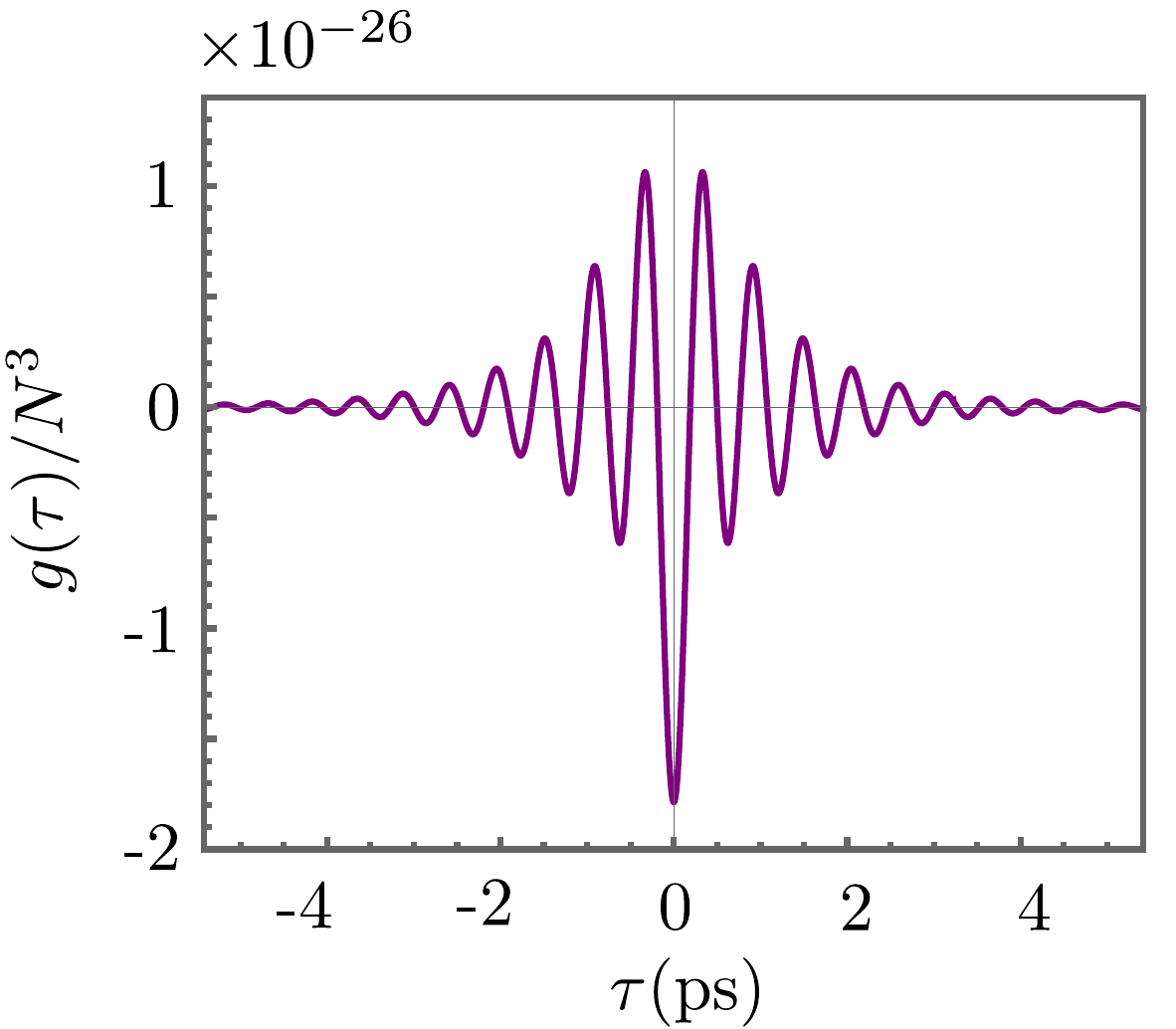}}
\quad
\subfigure{\includegraphics[width=7cm]{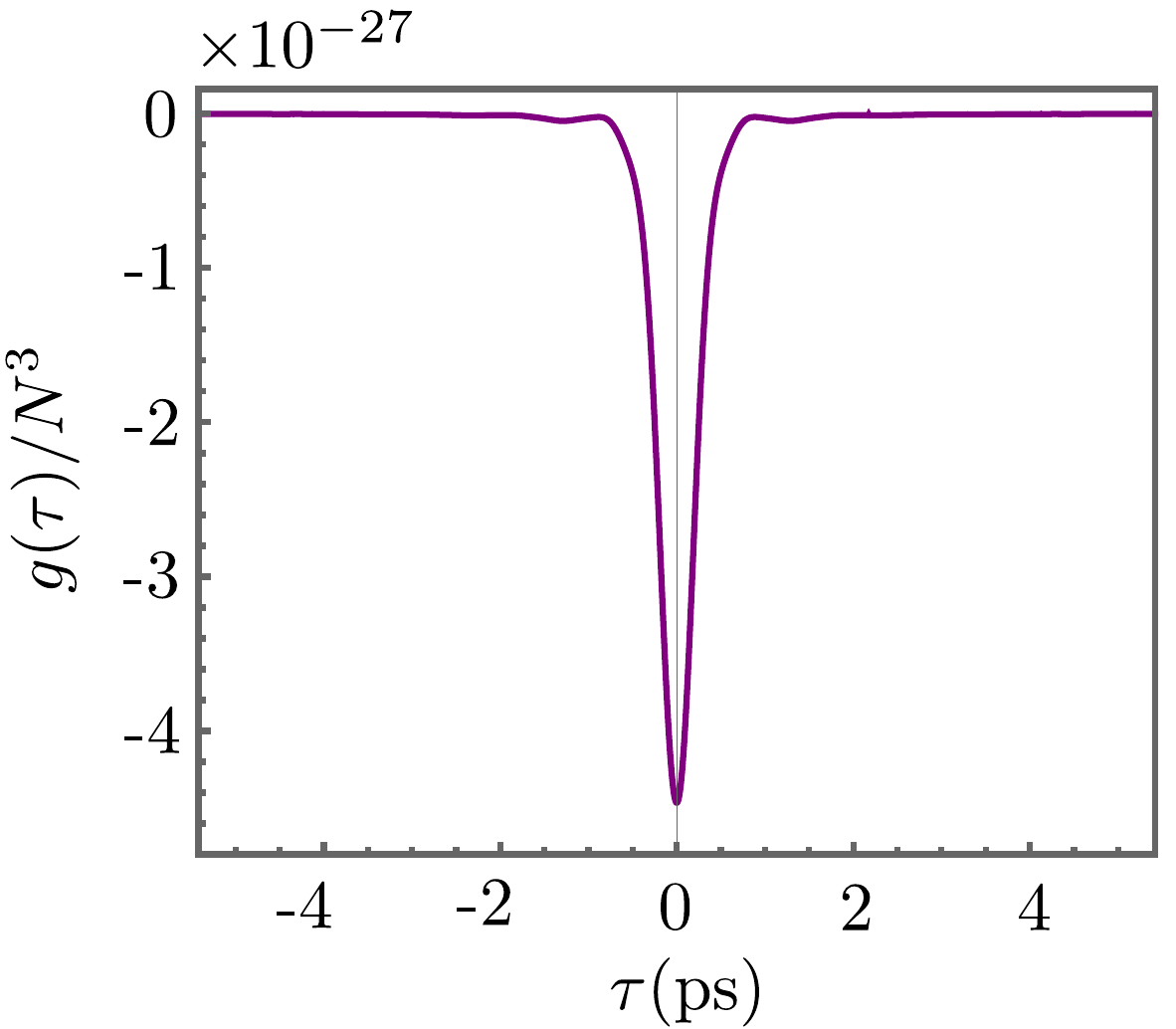}}
  \caption{Left: $V^{(a)}_{\{1,3\}}(\tau)/N^3$. Right: $V^{(b)}_{\{1,3\}}(\tau)/N^3$.}
  \label{V13_independentplots}
\end{figure}

The mixing between the main signal, $\hat{\mathcal{S}}^{(1)}[\delta \hat{E}_s(\Omega)]$, and $\hat{\mathcal{S}}^{(3)}[ \hat{E}^{(3)}_s(\Omega)]$ in the considered two-channel setup gives $\frac{1}{2} \langle \hat{\mathcal{S}}^{(3)}_\mathrm{ch1}\hat{\mathcal{S}}^{(1)}_\mathrm{ch2}+\hat{\mathcal{S}}^{(1)}_\mathrm{ch1}\hat{\mathcal{S}}^{(3)}_\mathrm{ch2}+\hat{\mathcal{S}}^{(3)}_\mathrm{ch2}\hat{\mathcal{S}}^{(1)}_\mathrm{ch1}+\hat{\mathcal{S}}^{(1)}_\mathrm{ch2}\hat{\mathcal{S}}^{(3)}_\mathrm{ch1}\rangle(\tau)\!=\! V^{(a)}_{\{1,3\}}(\tau) + V^{(b)}_{\{1,3\}}(\tau)$. Here, $V^{(a)}_{\{1,3\}}$ results from four equivalent processes: one in which the output of wave vector $\vec{k}_\Omega+\vec{k}'_\mathrm{ch1}-\vec{k}''_\mathrm{ch1}+\vec{k}'''_\mathrm{ch1}$ is multiplied with the output of vector $\vec{k}_{\Omega}+\vec{k}'_\mathrm{ch2}$, another in which the output of wave vector $\vec{k}_\Omega+\vec{k}'_\mathrm{ch2}-\vec{k}''_\mathrm{ch2}+\vec{k}'''_\mathrm{ch1}$ is multiplied with the output with $\vec{k}_{\Omega}+\vec{k}'_\mathrm{ch2}$, and the other two processes corresponding to swapped indices ch1 and ch2. $V^{(b)}_{\{1,3\}}$ originates from the product of field contributions with wave vectors $\vec{k}_\Omega+\vec{k}'_\mathrm{ch1}-\vec{k}''_\mathrm{ch2}+\vec{k}'''_\mathrm{ch2}$ and $\vec{k}_{\Omega}+\vec{k}'_\mathrm{ch2}$ and the corresponding index-swapped counterpart.
Explicitly we have
\begin{equation}
V^{(a)}_{\{1,3\}}(\tau)
=\frac{-C_2}{\sqrt{2}}
\sum_{t,s, u=\pm}\int^\infty_{-\infty} \!\!\!\!\!\mathrm{d}\Omega \!\int^\infty_0\!\!\!\!\!  \mathrm{d}\Omega' (\Omega\Omega') R^{(u)*}_{0} (\Omega', 0, 0)G^{(t,s)}_{00} (\Omega, \Omega',0,0, 0)[\cos(\Omega\tau)+\cos(\Omega'\tau)]\frac{n}{n_{\Omega'}}
\label{V13a}
\end{equation}
and
 \begin{equation}
V^{(b)}_{\{1,3\}}(\tau)
\frac{-C_2}{\sqrt{2}}
\sum_{t,s=\pm}\int^\infty_{-\infty} \!\!\!\!\!\mathrm{d}\Omega \!\int^\infty_0\!\!\!\!\!  \mathrm{d}\Omega' (\Omega\Omega') [R^{(s)*}_{0} (\Omega', 0, 0)G^{(t,t)}_{11} (\Omega, \Omega',\Omega',0, \tau)+R^{(s)*}_{0} (\Omega', 0, \tau)G^{(t,t)}_{22} (\Omega, \Omega',\Omega',0, \tau)]\frac{n}{n_{\Omega'}}.
\label{V13b}
\end{equation}
The behaviors of $V^{(a)}_{\{1,3\}}(\tau)$ and $V^{(b)}_{\{1,3\}}(\tau)$ can be seen in Fig.~\ref{V13_independentplots}.

\begin{figure}[h!]
    \centering
    \includegraphics[width=0.4\linewidth]{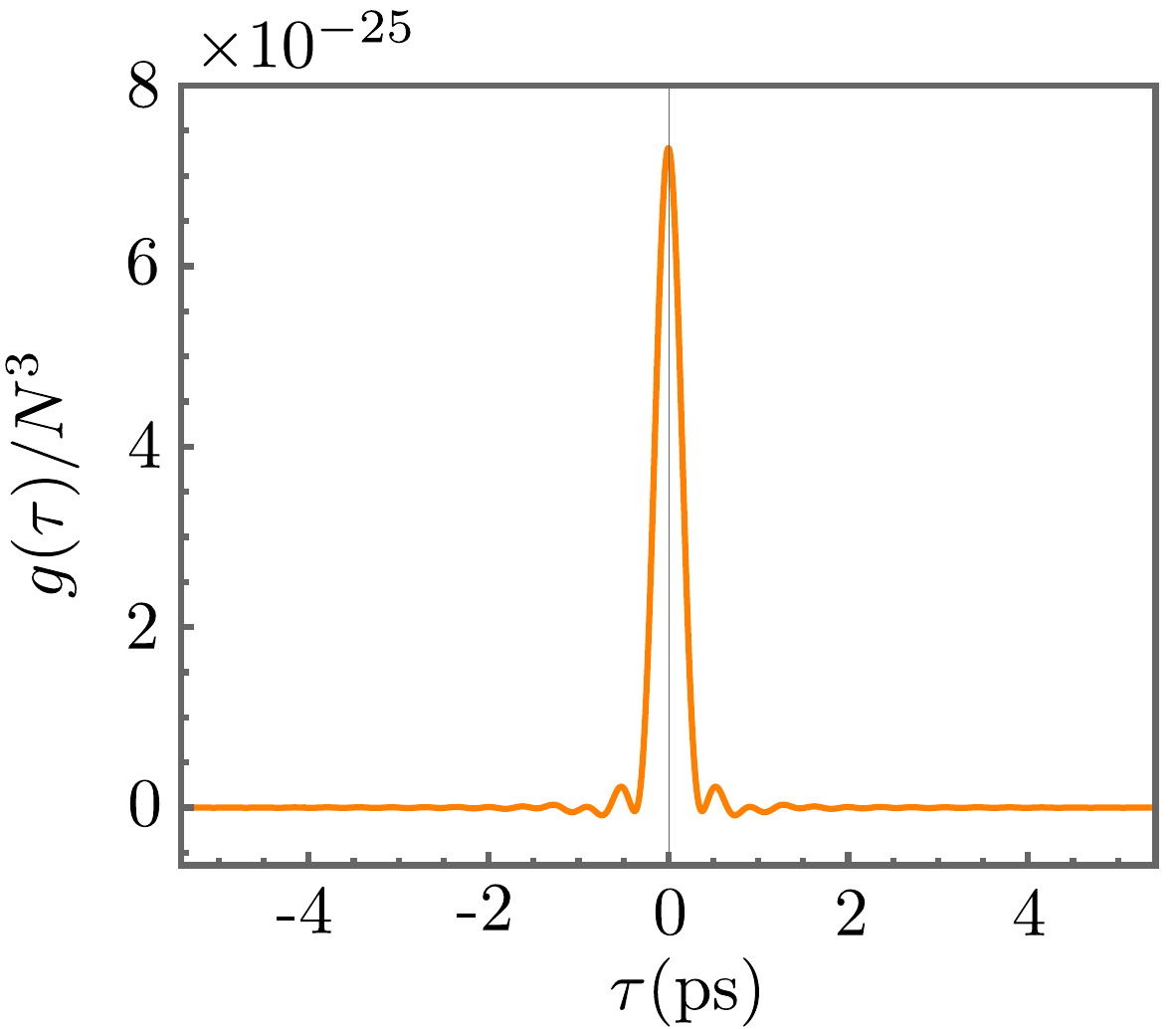}
 \caption{$V^{(c)}_{\{0,4\}}(\tau)/N^3$ profile.}
  \label{V04_independentplots}
\end{figure}

The last 4th-order contribution, from the crossterm between the base-SN signal and the signal $\hat{\mathcal{S}}^{(4)}[ \hat{E}^{(5)}(\omega)]$, has the form $ \frac{1}{2}\langle \hat{\mathcal{S}}^{(4)}_\mathrm{ch1}\hat{\mathcal{S}}^{(0)}_\mathrm{ch2}+\hat{\mathcal{S}}^{(0)}_\mathrm{ch1}\hat{\mathcal{S}}^{(4)}_\mathrm{ch2}+\hat{\mathcal{S}}^{(4)}_\mathrm{ch2}\hat{\mathcal{S}}^{(0)}_\mathrm{ch1}+\hat{\mathcal{S}}^{(0)}_\mathrm{ch2}\hat{\mathcal{S}}^{(4)}_\mathrm{ch1}\rangle(\tau)\!=\! V^{(a)}_{\{0,4\}}(\tau) + V^{(b)}_{\{0,4\}}(\tau) +V^{(c)}_{\{0,4\}}(\tau)$. As in the previous case, $V^{(a)}_{\{0,4\}}$ results from four equivalent processes: one in which the output of wave vector $\vec{k}_\mathrm{ch2}-\vec{k}'_\mathrm{ch2}+\vec{k}''_\mathrm{ch2}-\vec{k}'''_\mathrm{ch2}+\vec{k}''_\mathrm{ch1}$ is multiplied with the SN output of vector $\vec{k}_\mathrm{ch2}$, another in which the $\vec{k}_\mathrm{ch2}-\vec{k}'_\mathrm{ch2}+\vec{k}''_\mathrm{ch1}-\vec{k}'''_\mathrm{ch1}+\vec{k}''_\mathrm{ch1}$ contribution combines with the same SN output, and the processes with $\mathrm{ch1}$ and $\mathrm{ch2}$ labels swapped. $V^{(b)}_{\{0,4\}}$ originates from the product of field contributions with wave vectors $\vec{k}_\mathrm{ch2}-\vec{k}'_\mathrm{ch2}+\vec{k}''_\mathrm{ch1}-\vec{k}'''_\mathrm{ch2}+\vec{k}''_\mathrm{ch2}$ and $\vec{k}_\mathrm{ch2}$, while  $V^{(c)}_{\{0,4\}}$ results from the product of contributions with wave vectors $\vec{k}_\mathrm{ch2}-\vec{k}'_\mathrm{ch1}+\vec{k}''_\mathrm{ch1}-\vec{k}'''_\mathrm{ch2}+\vec{k}''_\mathrm{ch1}$ and $\vec{k}_\mathrm{ch2}$ (and the corresponding terms with swapped indices). The corresponding expressions are:
\begin{equation}
V^{(a)}_{\{0,4\}}(\tau)
=\frac{C_2}{\sqrt{5}}
\sum_{t,s=\pm}\int^\infty_{-\infty} \!\!\!\!\!\mathrm{d}\Omega \!\int^\infty_{-\infty}\!\!\!\!\!  \mathrm{d}\Omega' (\Omega\Omega') R^{(+)*}_{0} (\Omega', 0, 0)G^{(t,s)}_{00} (\Omega, \Omega',0,0, 0)[\cos(\Omega\tau)+\cos(\Omega'\tau)],
\label{V04a}
\end{equation}
 \begin{equation}
V^{(b)}_{\{0,4\}}(\tau)
=\frac{C_2}{\sqrt{5}}
\sum_{t=\pm}\int^\infty_{-\infty} \!\!\!\!\!\mathrm{d}\Omega \!\int^\infty_{-\infty}\!\!\!\!\!  \mathrm{d}\Omega' (\Omega\Omega') [R^{(+)*}_{0} (\Omega', 0, 0)G^{(t,t)}_{11} (\Omega, \Omega',\Omega',0, \tau)+R^{(+)*}_{0} (\Omega', 0, 0)G^{(t,t)}_{22} (\Omega, \Omega',\Omega',0, \tau)],
\label{V04b}
\end{equation}
and
 \begin{equation}
V^{(c)}_{\{0,4\}}(\tau)
=\frac{C_2}{\sqrt{5}}
\sum_{t=\pm}\int^\infty_{-\infty} \!\!\!\!\!\mathrm{d}\Omega \!\int^\infty_{-\infty}\!\!\!\!\!  \mathrm{d}\Omega' (\Omega\Omega') [R^{(+)*}_{1} (\Omega', \Omega, \tau)G^{(t,+)}_{01} (\Omega, \Omega',0,0, \tau)+R^{(+)*}_{2} (\Omega', \Omega, \tau)G^{(t,+)}_{02} (\Omega, \Omega',0,0, \tau)].
\label{V04c}
\end{equation}
Because both $V^{(a)}_{\{0,4\}}(\tau)$ and $V^{(b)}_{\{0,4\}}(\tau)$ turn out to vanish on the grounds of symmetry (for the same reason why Eq.~\eqref{signalcross4} vanishes), $V^{(c)}_{\{0,4\}}(\tau)$ is the only non-vanishing term in this series; its profile is shown in Fig.~\ref{V04_independentplots}.

 There is a minute difference between $N_{\mathrm{min}}$ in Fig.~\ref{photonnumber} and the corresponding $N$ at which the 4th-order contributions to the BA starts to dominate in the two-channel measurement, and it is related to the splitting of each single-channel contribution $\hat{E}^{(m)}_s$ into $2^{m-1}$ terms when two pulses are present.

\section{Phase matching, dispersion and absorption}

\begin{figure}[h!]
    \centering
    \includegraphics[width=0.4\linewidth]{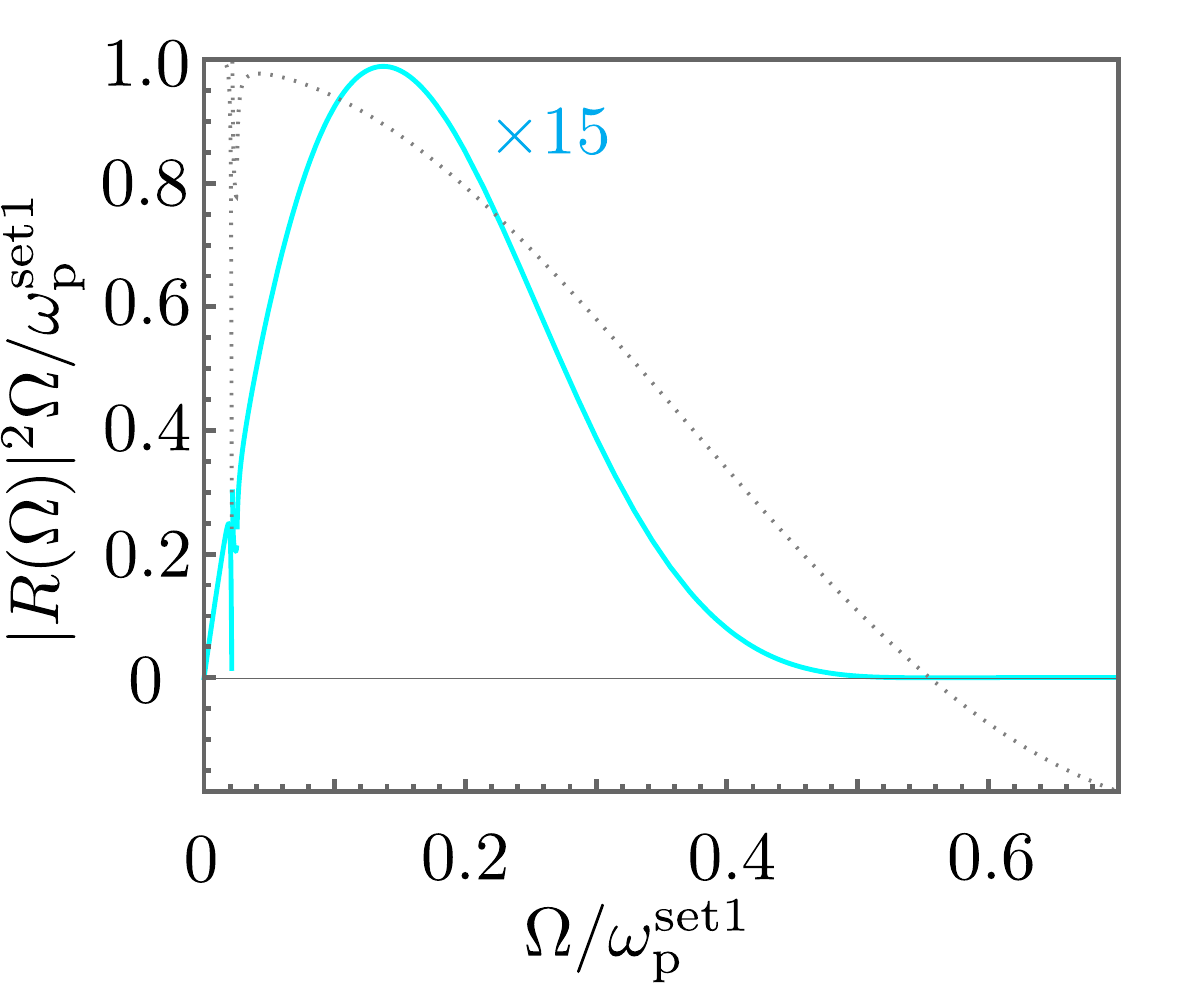}
 \caption{{(Full cyan) Gating function $|R(\Omega)|^2\Omega/\omega^{(\text{set}1)}_\mathrm{p}$ for the first set of parameters considered (scaled up by factor $15$). (Dotted gray) Corresponding phase-matching function $|\zeta_{\omega,\Omega}|/(d\frac{L \omega}{2c_0 n })$.}}
  \label{gatingA}
\end{figure}

For the first set of parameters, the RI in the MIR range up to 150 THz is modeled as~\cite{Leitenstorfer1999Sup}
\begin{equation}
n^{(\text{set}1)}_\Omega = 
 \Re \sqrt{6.7\left[1 + \frac{( 6.2 )^2 - ( 5.3)^2}{(5.3)^2 - \widetilde{\Omega}^2 - 0.09i|\widetilde{\Omega}|}\right]} \, ,
\end{equation}
where
$\widetilde{\Omega}=\Omega/(2\pi\times 10^{12} \text{THz})$. The corresponding gating function is shown in Fig.~\ref{gatingA}.
Since an absorption peak is found at around 5 THz, a cutoff is applied for frequencies below 18 THz when calculations involving the first set of parameters are considered.

Consideration of the second set of parameters requires the usage of a different RI profile due to the extra low frequency band selected by the gating function, partially overlapping with the absorption peak of the medium. In this case, the following dispersion profile is utilized~\cite{beneaSup}:
 \begin{equation}
n_\Omega 
=-0.0164\,\widetilde{\Omega}^6 + 
  0.1478\big|\widetilde{\Omega}\big|^5 -
  0.5185\,\widetilde{\Omega}^4 \\
  + 
  0.8974\big|\widetilde{\Omega}\big|^3 -
  0.7782\,\widetilde{\Omega}^2 + 
  0.3283\big|\widetilde{\Omega}\big| + 3.0657 \, .
\end{equation}
On top of that, an additional damping function is included in the gating function for the second parameter set in order to account for absorption. The absorption profile (extracted from Ref.~\cite{beneaSup}) is given by 
\begin{equation}
 \text{Abs}(\Omega)=\exp \left[-0.000618\,\widetilde{\Omega}^8 - 
   0.0000879\,\widetilde{\Omega}^6\right] \, .
   \label{Absorption}
\end{equation}

\begin{figure}[h!]
    \centering
    \includegraphics[width=0.4\linewidth]{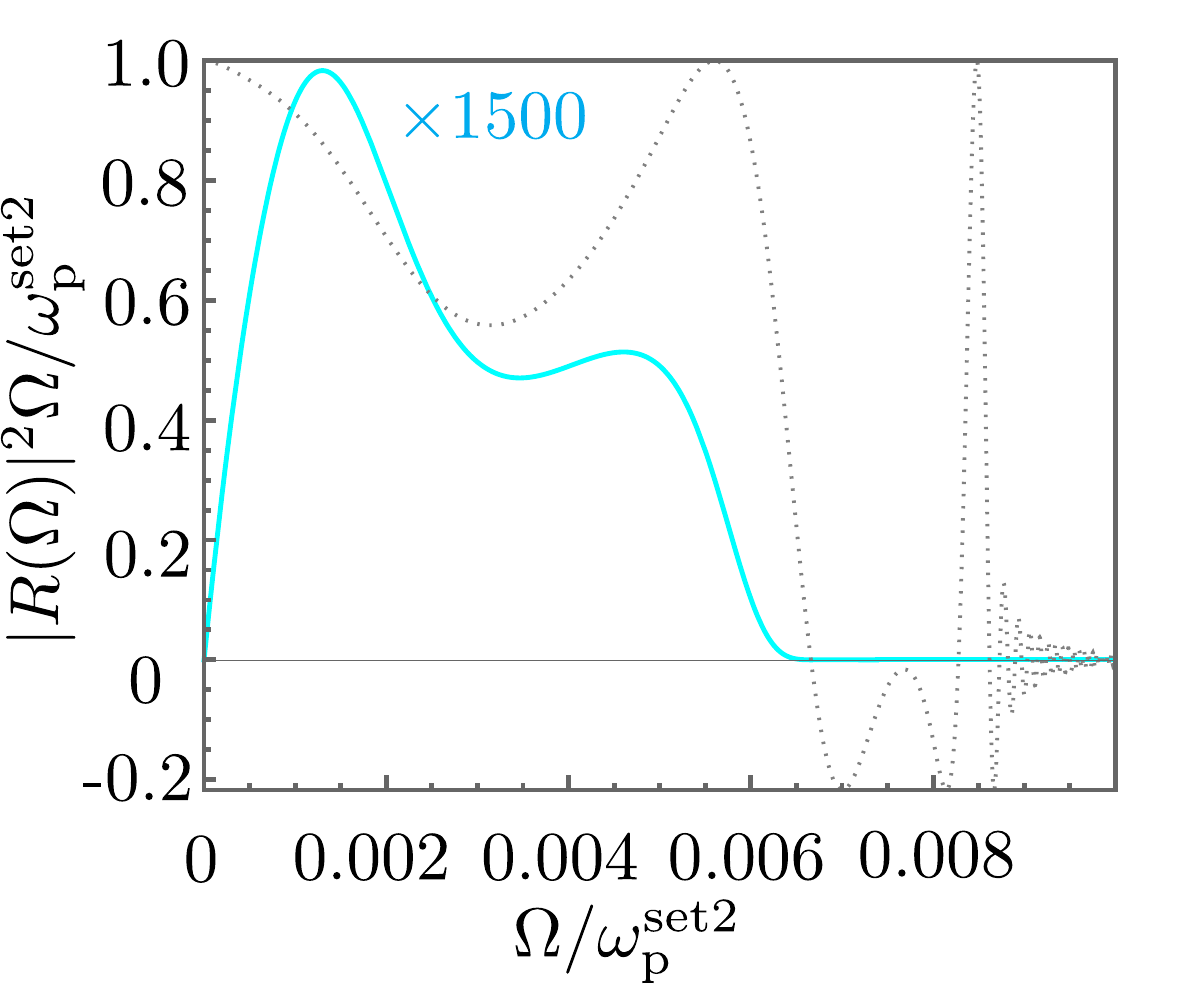}
 \caption{{(Full cyan) Gating function $ |R(\Omega)|^2 \text{Abs}(\Omega)\Omega/\omega^{(\text{set}2)}_\mathrm{p}$ for the second set of parameters considered (scaled up by 1500). (Dotted gray) Corresponding phase-matching function $|\zeta_{\omega,\Omega}|/(d\frac{L \omega}{2c_0 n })$.}}
  \label{gatingF}
\end{figure}

Fortunately, the effect from taking Eq.~\eqref{Absorption} into account is marginal, so that the validity of the paraxial quantization is not compromised. This also justifies the neglection of absorption when treating the two-channel experiment, in order to decrease computational effort. The resulting phase-matching function (including absorption) is shown in Fig.~\ref{gatingF}. 

\section{$\chi^{(3)}$ contribution}

An aspect that might arise when accounting for cascaded nonlinear processes is the possibility of four-wave mixing  taking place. The main contribution to the nonlinear polarization leading to the generation of such field components is proportional to the susceptibility $\chi^{(3)}$ and to the combination of terms $E^2_\mathrm{p}(t)\delta\hat{E}(t)$. Under the assumption of Kleinman symmetry~\cite{Boyd_bookSup}, there are six $\chi^{(3)}$ components that can generate measurable field components: $\chi^{(3)}_{xxzz}=\chi^{(3)}_{yyzz}$, $\chi^{(3)}_{xzxz}=\chi^{(3)}_{yzyz}$ and $\chi^{(3)}_{xzzx}=\chi^{(3)}_{yzzy}$. In fact, for a probe pulse propagating through a ZnTe crystal along the $[110]$ direction and polarized along the $[001]$ axis, the (third-order) nonlinear-polarization components generating additional electric fields are
\begin{equation}
\hat{P}^{(3)}_x(t)= \epsilon_0(\chi^{(3)}_{xxzz}+\chi^{(3)}_{xzxz}+\chi^{(3)}_{xzzx})E^2_\mathrm{p}(t)\delta\hat{E}_x(t)
\end{equation}
and
\begin{equation}
\hat{P}^{(3)}_y(t)= \epsilon_0(\chi^{(3)}_{yyzz}+\chi^{(3)}_{yzyz}+\chi^{(3)}_{yzzy})E^2_\mathrm{p}(t)\delta\hat{E}_y(t) \, .
\end{equation}
In terms of the basis spanned by $\vec{e}_z$, $\vec{e}_s$ and $\vec{e}_k$, this polarization has a longitudinal $\vec{e}_k$ component, which does not generate any outgoing field contribution, and a $\vec{e}_s$ component 
\begin{equation}
\hat{P}^{(3)}_s(t)= \epsilon_0(\chi^{(3)}_{xxzz}+\chi^{(3)}_{xzxz}+\chi^{(3)}_{xzzx})E^2_\mathrm{p}(t)\delta\hat{E}_s(t) \, ,
\end{equation}
with $\delta \hat{E}_s= \frac{1}{\sqrt{2}}(\delta \hat{E}_y-\delta \hat{E}_x)$. For brevity, we denote $X=\chi^{(3)}_{xxzz}+\chi^{(3)}_{xzxz}+\chi^{(3)}_{xzzx}$.
From our definition, the spectral decomposition of this expression gives
\begin{equation}
P^{(3)}_s(\omega)= \epsilon_0X\int^\infty_{-\infty}\hspace{-1.5mm}d\omega' I(\omega-\omega') \delta\hat{E}_s(\omega')e^{\frac{i r_k}{c_0}\left[\omega'n_{\omega'}+(\omega-\omega')n_{\omega-\omega'}-\omega n_\omega\right]} \, ,
\end{equation}
in which
\begin{equation}
I(\Omega)= 
\int^\infty_{-\infty}\hspace{-1.5mm}d\omega E^*_\mathrm{p}(\omega-\Omega) {E}_\mathrm{p}(\omega)e^{\frac{i r_k}{c_0}\left[\omega n_{\omega}-(\omega-\Omega)n_{\omega-\Omega}-\Omega n_\Omega\right]} \, .
\end{equation}
With this polarization, we find the following approximate solution to the paraxial equation:
\begin{equation}
\hat{E}^{\chi^{(3)}}_s(\omega)\hspace{-0.5mm}=\hspace{-0.5mm} \frac{XL\omega}{2nc_0}\hspace{-1mm}\int^\infty_{-\infty}\hspace{-3.5mm}d\omega' \mathrm{d}\omega'' E^*_\mathrm{p}(\omega'+\omega''-\omega) E_\mathrm{p}(\omega') \delta\hat{E}_s(\omega'')\hspace{0.3mm}\text{sinc}\hspace{-0.45mm}\left\{\frac{L}{2c_0}\left[\omega''n_{\omega''}-(\omega'+\omega''-\omega)n_{\omega'+\omega''-\omega}-\omega n_\omega+\omega'n_{\omega'}\right]\right\} .\label{Efield_chi3_sup}
\end{equation}

When all the 3 frequencies in Eq.~\eqref{Efield_chi3_sup} are in the NIR, $\text{sinc}\hspace{-0.45mm}\left\{\frac{L}{2c_0}\left[\omega''n_{\omega''}-(\omega'+\omega''-\omega)n_{\omega'+\omega''-\omega}-\omega n_\omega+\omega'n_{\omega'}\right]\right\}\approx 1$. This allows one to write the signal operator resulting after the ellipsometry step as (note that even though the spatial argument of the electric field has been omitted, the effect of the beam waist is accounted for in the value of $A^{(3)}$ used for the signal operator)
\begin{equation}
\hat{\mathcal{S}}^{\chi^{(3)}}= \frac{4\sqrt{\epsilon_0}XL\kappa i}{w^2_0\sqrt{3\pi\hbar c_0 n}}\int^\infty_0 \hspace{-1.5mm} d\omega\int^\infty_{-\infty}\hspace{-1.5mm}d\omega' \sqrt{\omega'}\sqrt{\frac{n}{n_{\omega'}}}F(\omega'-\omega) \alpha^*_\mathrm{p}(\omega) \hat{a}(\omega') + \text{H.c.} \, .
\label{signalchi3}
\end{equation}

Eq.~\eqref{signalchi3} does not give any contributions at 2nd order, since its crossterm with $\hat{\mathcal{S}}^{(0)}$ vanishes. At 4th order, only the square of Eq.~\eqref{signalchi3} contributes and gives
\begin{equation}
\langle[\hat{\mathcal{S}}^{\chi^{(3)}}]^2\rangle= \frac{4N^3}{3}\left(\frac{L \omega_\mathrm{p}}{nc_0}X\right)^{2} \left(\frac{ \hbar }{4\pi^2 \epsilon_0 c_0 n w^2_0}\right)^2\int^\infty_0 \hspace{-1.5mm} d\omega\hspace{-0.6mm}\int^\infty_{0}\hspace{-1.6mm}d\omega'\hspace{-0.6mm}\int^\infty_{0}\hspace{-1.6mm}d\omega'' \frac{n}{n_{\omega'}}\omega'F(\omega'-\omega)F^*(\omega'-\omega'') \frac{\alpha^*_\mathrm{p}(\omega)\alpha_\mathrm{p}(\omega'')}{\beta} \, . \label{thatonevariance}
\end{equation}

 Precise values of the relevant tensor components of $\chi^{(3)}$ cannot, to the best of our knowledge, be found reliably in the literature. Indirect $\chi^{(3)}$ measurements based on the RI modulation through Kerr effect lead to a wide span of reported values~\cite{chi3Sup, chi3_ASup, chi3_BSup, Denis_supSup}. The large fluctuations in the found values for linear combinations of $\chi^{(3)}$ tensor components can be traced back to the sampled frequency ranges, with measurements near the 2-photon-absorption frequency (when the sum of the frequencies of two photons matches the band gap of the NX) leading to resonantly enhanced values of the 3rd-order nonlinear susceptibility. At such frequencies, undesired effects like crystal heating (through generation of electron-hole pairs) and probe depletion might take place. For ZnTe, used in Ref.~\cite{beneaSup}, the band gap corresponds to a frequency of $\Delta\omega\approx 550 \text{THz}$, while for AgGaS$_2$, used in Ref.~\cite{vacuum_sampSup}, the value is $\Delta\omega\approx 660 \text{THz}$. Adopting the $\chi^{(3)}$ value extracted from Ref.~\cite{chi3_BSup} and assuming all tensor components in $X$ to be the same, one arrives at $\langle[\hat{\mathcal{S}}^{\chi^{(3)}}]^2\rangle\sim 10^{-21} N^3$ for the 1st set of parameters. $\langle[\hat{\mathcal{S}}^{\chi^{(3)}}]^2\rangle$ might be as large as (or possibly larger than) the signal-variance contributions originating from cascaded $\chi^{(2)}$ processes (depending on the choice of NX and probe frequency range). The contribution described by Eq.~\eqref{thatonevariance}, however, relates to a process of nature similar to self-phase modulation and therefore affects solely the electric-field fluctuations in the probe pulse (SN enhancement). These $\chi^{(3)}$ contributions have no connections to the sampled MIR vacuum and can be suppressed in a measurement (e.g., by investigating and avoiding the presence of self-phase modulation in the spectrum of the probe pulse), therefore justifying their neglection in our Letter.

\section{Quantum-state evolution}

The quantum states participating in EO sampling involve two polarizations and an ultrabroadband continuous range of frequencies.
The treatment of a multimode problem can be circumvented by the use of non-monochromatic modes covering the involved NIR and the MIR frequency ranges, leading to an effective description of the probe-driven interaction as a two-mode squeezing between the $s$-polarized MIR and NIR modes~\cite{ShoSup}. In the subcycle sampling regime, the corresponding non-monochromatic mode operators exhibit some unusual properties that can be related to the presence of virtual particles. As an example, the MIR annihilation operator does not completely annihilate the vacuum state $|\{0\}_\Omega\rangle$ defined over the continuous set of the MIR frequencies. For the sake of simplicity, we shall ignore these properties. If we additionally adopt a fully quantum description of the probe,  the state evolution can then be treated in terms of an effective 3-mode interaction, with corresponding operators $\hat{a}_\mathrm{MIR}$, $\hat{a}_\mathrm{NIR}$ and $\hat{a}_\mathrm{p}$ for the $s$-polarized MIR, $s$-polarized NIR and $z$-polarized NIR (probe), respectively. One can then
enforce energy conservation on the Hamiltonian (neglecting time-ordering effects)
and exclude up-conversion to frequencies above the NIR (as well as the
inverse process), so that the simplified action $\hat{\mathrm{S}}=\int \mathrm{d}t \hat{H}$ becomes
\begin{equation}
\hat{\mathrm{S}}=A\hat{a}^\dagger_\mathrm{MIR}\hat{a}^\dagger_\mathrm{NIR}\hat{a}_\mathrm{p}+A^*\hat{a}_\mathrm{MIR}\hat{a}_\mathrm{NIR}\hat{a}^\dagger_\mathrm{p}+C\hat{a}_\mathrm{MIR}\hat{a}^\dagger_\mathrm{NIR}\hat{a}_\mathrm{p}+C^*\hat{a}^\dagger_\mathrm{MIR}\hat{a}_\mathrm{NIR}\hat{a}^\dagger_\mathrm{p}
\end{equation}
and the unitary evolution operator reads
\begin{equation}
\hat{U}=\exp \left\{\mathcal{A}\hat{a}^\dagger_\mathrm{MIR}\hat{a}^\dagger_\mathrm{NIR}\hat{a}_\mathrm{p}-\mathcal{A}^*\hat{a}_\mathrm{MIR}\hat{a}_\mathrm{NIR}\hat{a}^\dagger_\mathrm{p}+\mathcal{C}\hat{a}_\mathrm{MIR}\hat{a}^\dagger_\mathrm{NIR}\hat{a}_\mathrm{p}-\mathcal{C}^*\hat{a}^\dagger_\mathrm{MIR}\hat{a}_\mathrm{NIR}\hat{a}^\dagger_\mathrm{p}\right\}  ,
\label{evolutionU}
\end{equation}
with $\mathcal{A}=-iA$ and $\mathcal{C}=-iC$ being coefficients depending on the field quantization considered and on the properties, geometry and modeling of the NX (so that $\mathcal{A}$ and $\mathcal{C}$ are proportional to linear combinations of elements of the nonlinear susceptibility). If for all 3 considered modes neither annihilation nor creation is favored relative to each other (what could happen, e.g., by considering frequencies close to the NX resonance), $|A|=|C|$ holds.

For the EO sampling, we consider the input state $|\alpha, 0, 0\rangle$, where the 1st, 2nd and 3rd entries stand for the probe ($z$-polarized NIR mode), MIR and $s$-polarized NIR modes, respectively. The probe is initially in the coherent state of amplitude $\alpha$ generated by a corresponding displacement operator, while both the MIR and $s$-polarized NIR modes are in their vacuum states. Through Taylor expansion, the first-order correction to the initial state caused by Eq.~\eqref{evolutionU} is given by
\begin{equation}
|\text{out}_1\rangle=\text{ln} \hat{U}|\alpha, 0, 0\rangle = \mathcal{A}\alpha |\alpha, 1, 1\rangle \, , \label{out_state_1}
\end{equation}
while the second-order one reads
\begin{equation}
|\text{out}_2\rangle=\frac{1}{2}\text{ln}^2 \hat{U}|\alpha, 0, 0\rangle = \Big( \mathcal{A}^2\alpha^2|\alpha, 2, 2\rangle -\frac{1}{2}|\mathcal{A}|^2\alpha \hat{a}^\dagger_\mathrm{p} |\alpha, 0, 0\rangle+\frac{1}{\sqrt{2}}\mathcal{A}\mathcal{C}\alpha^2|\alpha, 0, 2\rangle-\frac{1}{\sqrt{2}}\mathcal{A}\mathcal{C}^*\alpha\hat{a}^\dagger_\mathrm{p}|\alpha, 2, 0\rangle\Big) \, .\label{out_state_2}
\end{equation}
The conjunction of quarter-wave plate, Wollaston prism and photon detection leads to the signal operator
\begin{equation}
\hat{\mathcal{S}}=i\big(\hat{a}^\dagger_\mathrm{p}\hat{a}_\mathrm{NIR}-\hat{a}^\dagger_\mathrm{NIR}\hat{a}_\mathrm{p}\big) \, , \label{simple_signal}
\end{equation}
which can be applied on \eqref{out_state_1} and \eqref{out_state_2} to give
\begin{equation}
|\text{sig}_1\rangle= \hat{\mathcal{S}}|\text{out}_1\rangle = i\mathcal{A}\alpha\Big( \hat{a}^\dagger_\mathrm{p}|\alpha, 1, 0\rangle -\sqrt{2}\alpha |\alpha, 1, 2\rangle\Big)  \, , \label{sig_state_1}
\end{equation}
and
\begin{align}
|\text{sig}_2\rangle= \hat{\mathcal{S}}|\text{out}_2\rangle =&  i\Big[\sqrt{2} \mathcal{A}^2\alpha^2\hat{a}^\dagger_\mathrm{p}|\alpha, 2, 1\rangle -\sqrt{3} \mathcal{A}^2\alpha^3|\alpha, 2, 3\rangle+\frac{1}{2}|\mathcal{A}|^2\alpha(\hat{a}^\dagger_\mathrm{p}\alpha+1) |\alpha, 0, 1\rangle+\mathcal{A}\mathcal{C}\alpha^2\hat{a}^\dagger_\mathrm{p}|\alpha, 0, 1\rangle\nonumber\\
&-\sqrt{\frac{3}{2}}\mathcal{A}\mathcal{C}\alpha^3|\alpha, 0, 3\rangle+\frac{1}{\sqrt{2}}\mathcal{A}\mathcal{C}^*\alpha(\hat{a}^\dagger_\mathrm{p}\alpha+1)|\alpha, 2, 1\rangle\Big]  \, .\label{sig_state_2}
\end{align}
Together with
\begin{equation}
|\text{sig}_0\rangle = \hat{\mathcal{S}}|\alpha,0,0\rangle = -i\alpha|\alpha, 0, 1\rangle \, ,
\end{equation}
which gives the (shot-noise) contribution $\langle\text{sig}_0|\text{sig}_0\rangle=|\alpha|^2$, and $\langle \hat{\mathcal{S}}\rangle =0$,
the expectation value for the signal variance up to the second order is given by $\langle \hat{\mathcal{S}}^2\rangle =\langle\text{sig}_0|\text{sig}_0\rangle+ \langle \text{sig}_1|\text{sig}_1\rangle+\langle \text{sig}_0|\text{sig}_2\rangle+\langle \text{sig}_2|\text{sig}_0\rangle$, with
\begin{equation}
\langle \text{sig}_1|\text{sig}_1\rangle = |\mathcal{A}|^2|\alpha|^2(1+3|\alpha|^2) \label{sigsig1}
\end{equation}
and
\begin{equation}
\langle \text{sig}_0|\text{sig}_2\rangle+\langle \text{sig}_2|\text{sig}_0\rangle=-|\mathcal{A}|^2|\alpha|^2-(|\mathcal{A}|^2+\mathcal{A}\mathcal{C}+\mathcal{A}^*\mathcal{C}^*)|\alpha|^4 \, .
\label{sigsig2}
\end{equation}
In other words, the total signal variance takes the form
\begin{equation}
\langle \hat{\mathcal{S}}^2\rangle =|\alpha|^2+ (2|\mathcal{A}|^2-\mathcal{A}\mathcal{C}-\mathcal{A}^*\mathcal{C}^*)|\alpha|^4 \, ,
\end{equation}
where the EO term is proportional to $|\alpha|^4\propto N^2$. Note that $|\text{out}_1\rangle$ (which has the same MIR state as $|\text{sig}_1\rangle$) contains a MIR photon, while $|\text{out}_2\rangle$ contains contributions from states both with and without MIR photons, but only the latter contribute to the signal at second order. It is insightful to reconnect this state-evolution approach to the Heisenberg-picture calculations presented in this Letter. One can notice that  $\langle \text{sig}_1|\text{sig}_1\rangle$ and $\langle \text{sig}_0|\text{sig}_2\rangle+\langle \text{sig}_2|\text{sig}_0\rangle$ are related to both $\langle \alpha, 0,0| [\hat{\mathcal{S}}^{(1)}]^2|\alpha, 0, 0\rangle=\langle [\hat{\mathcal{S}}^{(1)}]^2\rangle$ [the superscript (1) here denotes first-order operator evolution, as in the main text] and $\langle \hat{\mathcal{S}}^{(2)}\hat{\mathcal{S}}^{(0)}\rangle+\langle \hat{\mathcal{S}}^{(0)}\hat{\mathcal{S}}^{(2)}\rangle$, since
\begin{equation}
\langle [\hat{\mathcal{S}}^{(1)}]^2\rangle=\langle [\hat{\mathcal{S}},\text{ln}\hat{U}]^2\rangle =-\langle \text{ln}\hat{U}\hat{\mathcal{S}}^2\text{ln}\hat{U}\rangle +\langle [\text{ln}\hat{U} \hat{\mathcal{S}}]^2\rangle -\langle \hat{\mathcal{S}}\text{ln}^2\hat{U}\hat{\mathcal{S}}\rangle +\langle [\hat{\mathcal{S}}\text{ln}\hat{U}]^2\rangle \label{evolutions_dissection1} 
\end{equation}
and
\begin{align}
\langle \hat{\mathcal{S}}^{(2)}\hat{\mathcal{S}}^{(0)}\rangle+\langle \hat{\mathcal{S}}^{(0)}\hat{\mathcal{S}}^{(2)}\rangle=&
\frac{1}{2}\langle [[\hat{\mathcal{S}},\text{ln}\hat{U}],\text{ln}\hat{U}]\hat{\mathcal{S}}\rangle+\frac{1}{2}\langle \hat{\mathcal{S}}[[\hat{\mathcal{S}},\text{ln}\hat{U}],\text{ln}\hat{U}]\rangle\nonumber\\
&=\frac{1}{2}\langle \text{ln}^2\hat{U}\hat{\mathcal{S}}^2\rangle+\frac{1}{2}\langle \hat{\mathcal{S}}^2\text{ln}^2\hat{U}\rangle -\langle [\text{ln}\hat{U} \hat{\mathcal{S}}]^2\rangle +\langle \hat{\mathcal{S}}\text{ln}^2\hat{U}\hat{\mathcal{S}}\rangle -\langle [\hat{\mathcal{S}}\text{ln}\hat{U}]^2\rangle  \, .\label{evolution_dissection2}
\end{align}
When combining Eqs.~\eqref{evolutions_dissection1} and \eqref{evolution_dissection2}, the terms that do not lead to either $\langle \text{sig}_1|\text{sig}_1\rangle$ or $\langle \text{sig}_0|\text{sig}_2\rangle+\langle \text{sig}_2|\text{sig}_0\rangle$ mutually cancel. Note that $\text{ln}\hat{U}$ is anti-Hermitian, so that $\langle \text{out}_1|=-\langle \alpha, 0 ,0|\text{ln}\hat{U}$ and therefore
$\langle[ \hat{\mathcal{S}}^{(1)}]^2\rangle+\langle \hat{\mathcal{S}}^{(2)}\hat{\mathcal{S}}^{(0)}\rangle+\langle \hat{\mathcal{S}}^{(0)}\hat{\mathcal{S}}^{(2)}\rangle=\langle \text{sig}_1|\text{sig}_1\rangle+\langle \text{sig}_0|\text{sig}_2\rangle+\langle \text{sig}_2|\text{sig}_0\rangle$. The terms that mutually cancel, however, are still present in the evolved signals and result in contributions to the signal variance arising from $\hat{\mathcal{S}}^{(1)}$ and $\hat{\mathcal{S}}^{(2)}$ that carry contributions from both $|\text{sig}_1\rangle$ and $|\text{sig}_2\rangle$.

Explicit evolution of the signal operator \eqref{simple_signal} results in
\begin{equation}
\hat{\mathcal{S}}^{(1)}=-i[\hat{n}_\mathrm{p}-\hat{n}_\mathrm{NIR}][(\mathcal{A}-\mathcal{C}^*)\hat{a}^\dagger_\mathrm{MIR}-(\mathcal{A}^*-\mathcal{C})\hat{a}_\mathrm{MIR}]
\end{equation}
and
\begin{align}
\hat{\mathcal{S}}^{(2)}&=\frac{i}{2}(|\mathcal{A}|^2-|\mathcal{C}|^2)[\hat{n}_\mathrm{p}-\hat{n}_\mathrm{NIR}][\hat{a}^\dagger_\mathrm{p}\hat{a}_\mathrm{NIR}-\hat{a}^\dagger_\mathrm{NIR}\hat{a}_\mathrm{p}]\nonumber\\
&-i[\mathcal{A}\hat{a}^\dagger_\mathrm{MIR}\hat{a}^\dagger_\mathrm{NIR}\hat{a}_\mathrm{p}+\mathcal{A}^*\hat{a}_\mathrm{MIR}\hat{a}_\mathrm{NIR}\hat{a}^\dagger_\mathrm{p}+\mathcal{C}\hat{a}_\mathrm{MIR}\hat{a}^\dagger_\mathrm{NIR}\hat{a}_\mathrm{p}+\mathcal{C}^*\hat{a}^\dagger_\mathrm{MIR}\hat{a}_\mathrm{NIR}\hat{a}^\dagger_\mathrm{p}][(\mathcal{A}-\mathcal{C}^*)\hat{a}^\dagger_\mathrm{MIR}-(\mathcal{A}^*-\mathcal{C})\hat{a}_\mathrm{MIR}]\, .
\label{long_signal2}
\end{align}
It is worth mentioning that, by considering in this Letter only nested convolutions linear in the unperturbed quantum fields $\delta\hat{E}$, we are keeping solely the terms proportional to $\hat{n}_\mathrm{p}=\hat{a}_p^\dagger \hat{a}_p$ in the above equations. These are the terms that give the $|\alpha|^4$ contributions to the signal variances. 

Using Eq.~\eqref{long_signal2}, one can calculate
\begin{align}
\hat{\mathcal{S}}^{(0)}\hat{\mathcal{S}}^{(2)}+\hat{\mathcal{S}}^{(2)}\hat{\mathcal{S}}^{(0)}&=-(|\mathcal{A}|^2-|\mathcal{C}|^2)\Big\{\hat{n}_\mathrm{p}(\hat{n}_\mathrm{NIR}+1)-\hat{n}_\mathrm{NIR}(\hat{n}_\mathrm{p}+1)-(\hat{a}^\dagger_\mathrm{p})^2\hat{a}^2_\mathrm{NIR}+(\hat{a}^\dagger_\mathrm{NIR})^2\hat{a}^2_\mathrm{p}\nonumber\\
&[\hat{n}_\mathrm{p}-\hat{n}_\mathrm{NIR}][-\hat{n}_\mathrm{p}(\hat{n}_\mathrm{NIR}+1)-\hat{n}_\mathrm{NIR}(\hat{n}_\mathrm{p}+1)+(\hat{a}^\dagger_\mathrm{p})^2\hat{a}^2_\mathrm{NIR}-(\hat{a}^\dagger_\mathrm{NIR})^2\hat{a}^2_\mathrm{p}]\Big\}\nonumber\\
&+\Big\{2[\mathcal{A}\hat{a}^\dagger_\mathrm{MIR}\hat{a}^\dagger_\mathrm{NIR}\hat{a}_\mathrm{p}+\mathcal{A}^*\hat{a}_\mathrm{MIR}\hat{a}_\mathrm{NIR}\hat{a}^\dagger_\mathrm{p}+\mathcal{C}\hat{a}_\mathrm{MIR}\hat{a}^\dagger_\mathrm{NIR}\hat{a}_\mathrm{p}+\mathcal{C}^*\hat{a}^\dagger_\mathrm{MIR}\hat{a}_\mathrm{NIR}\hat{a}^\dagger_\mathrm{p}](\hat{a}^\dagger_\mathrm{p}\hat{a}_\mathrm{NIR}-\hat{a}^\dagger_\mathrm{NIR}\hat{a}_\mathrm{p})\nonumber\\
&+(\hat{n}_\mathrm{p}-\hat{n}_\mathrm{NIR})[(\mathcal{A}+\mathcal{C}^*)\hat{a}^\dagger_\mathrm{MIR}+(\mathcal{A}^*+\mathcal{C})\hat{a}_\mathrm{MIR}]\Big\}[(\mathcal{A}-\mathcal{C}^*)\hat{a}^\dagger_\mathrm{MIR}-(\mathcal{A}^*-\mathcal{C})\hat{a}_\mathrm{MIR}]\, ,
\label{long_signal02}
\end{align}
which leads to
\begin{equation}
\langle \hat{\mathcal{S}}^{(2)}\hat{\mathcal{S}}^{(0)}\rangle+\langle \hat{\mathcal{S}}^{(0)}\hat{\mathcal{S}}^{(2)}\rangle = -(|\mathcal{A}|^2+|\mathcal{C}|^2-\mathcal{A}\mathcal{C}-\mathcal{A}^*\mathcal{C}^*)|\alpha|^2+(|\mathcal{A}|^2-|\mathcal{C}|^2)|\alpha|^4\, .
\end{equation}
Similarly,
\begin{equation}
\langle [\hat{\mathcal{S}}^{(1)}]^2\rangle= (|\mathcal{A}|^2+|\mathcal{C}|^2-\mathcal{A}\mathcal{C}-\mathcal{A}^*\mathcal{C}^*)(|\alpha|^2+|\alpha|^4)\, .
\end{equation}
It is clear that the above equations are linear combinations of \eqref{sigsig1} and \eqref{sigsig2}, emphasizing the fact that both $\langle \hat{\mathcal{S}}^{(2)}\hat{\mathcal{S}}^{(0)}\rangle+\langle \hat{\mathcal{S}}^{(0)}\hat{\mathcal{S}}^{(2)}\rangle$ and $\langle [\hat{\mathcal{S}}^{(1)}]^2\rangle$ contain contributions from populated MIR states. When the relation $|A|=|C|$ holds, the $|\alpha|^4$-dependent term in $\langle \hat{\mathcal{S}}^{(2)}\hat{\mathcal{S}}^{(0)}\rangle+\langle \hat{\mathcal{S}}^{(0)}\hat{\mathcal{S}}^{(2)}\rangle$ vanishes, in agreement with Eq.~\eqref{signalcross_ch1}.

For a two-channel setup, similar calculations can be performed using the initial state $|\alpha_\mathrm{ch1}, \alpha_\mathrm{ch2},0_\mathrm{MIR},0_\mathrm{NIR,ch1},0_\mathrm{NIR,ch2}\rangle$. The signal operator has to be replaced by two operators: $\hat{\mathcal{S}}_\mathrm{ch1}=i(\hat{a}^\dagger_\mathrm{p, ch1}\hat{a}_\mathrm{NIR, ch1}-\hat{a}^\dagger_\mathrm{NIR, ch1}\hat{a}_\mathrm{p, ch1})$ and $\hat{\mathcal{S}}_\mathrm{ch2}=i(\hat{a}^\dagger_\mathrm{p, ch2}\hat{a}_\mathrm{NIR, ch2}-\hat{a}^\dagger_\mathrm{NIR, ch2}\hat{a}_\mathrm{p, ch2})$. Analogously, in the evolution operator in place of  $\mathcal{A}\hat{a}_\mathrm{p}\hat{a}^\dagger_\mathrm{NIR}\hat{a}^\dagger_\mathrm{MIR}$ one should consider $\mathcal{A}_\mathrm{ch1}\hat{a}_\mathrm{p, ch1} \hat{a}^\dagger_\mathrm{NIR, ch1}\hat{a}^\dagger_\mathrm{MIR}+\mathcal{A}_\mathrm{ch2}\hat{a}_\mathrm{p, ch2}\hat{a}^\dagger_\mathrm{NIR, ch2}\hat{a}^\dagger_\mathrm{MIR}$ [$\hat{a}_\mathrm{MIR}=(\hat{a}_\mathrm{MIR, ch1}+i\hat{a}_\mathrm{MIR,ch2})/\sqrt{2}$], with similar replacements for the other terms. This can be clearly seen, e.g., for the first-order state
\begin{equation}
|\text{out}'_1\rangle = \mathcal{A}_\mathrm{ch1}\alpha_\mathrm{ch1}|\alpha_\mathrm{ch1},\alpha_\mathrm{ch2},1,1,0\rangle+\mathcal{A}_\mathrm{ch2}\alpha_\mathrm{ch2}|\alpha_\mathrm{ch1},\alpha_\mathrm{ch2},1,0,1\rangle \, ,
\end{equation}
which can generate two contributions
\begin{equation}
|\text{sig}'_{1,\mathrm{ch1}}\rangle = i\mathcal{A}_\mathrm{ch1}\alpha_\mathrm{ch1}\hat{a}^\dagger_\mathrm{p,ch1}|\alpha_\mathrm{ch1},\alpha_\mathrm{ch2},1,0,0\rangle-i\sqrt{2}\mathcal{A}_\mathrm{ch1}\alpha^2_\mathrm{ch1}|\alpha_\mathrm{ch1},\alpha_\mathrm{ch2},1,2,0\rangle-i\mathcal{A}_\mathrm{ch2}\alpha_\mathrm{ch1}\alpha_\mathrm{ch2}|\alpha_\mathrm{ch1},\alpha_\mathrm{ch2},1,1,1\rangle \, ,
\end{equation}
and
\begin{equation}
|\text{sig}'_{1,\mathrm{ch2}}\rangle = i\mathcal{A}_\mathrm{ch2}\alpha_\mathrm{ch2}\hat{a}^\dagger_\mathrm{p,ch2}|\alpha_\mathrm{ch1},\alpha_\mathrm{ch2},1,0,0\rangle-i\sqrt{2}\mathcal{A}_\mathrm{ch2}\alpha^2_\mathrm{ch2}|\alpha_\mathrm{ch1},\alpha_\mathrm{ch2},1,0,2\rangle-i\mathcal{A}_\mathrm{ch1}\alpha_\mathrm{ch1}\alpha_\mathrm{ch2}|\alpha_\mathrm{ch1},\alpha_\mathrm{ch2},1,1,1\rangle \, .
\end{equation}
The cross-signal contribution to the variance (relevant for the corresponding experiment) then reads
\begin{equation}
\langle\text{sig}'_{1,\mathrm{ch1}}|\text{sig}'_{1,\mathrm{ch2}}\rangle+\langle\text{sig}'_{1,\mathrm{ch2}}|\text{sig}'_{1,\mathrm{ch1}}\rangle = 2(\mathcal{A}^*_\mathrm{ch1}\mathcal{A}_\mathrm{ch2}+\mathcal{A}_\mathrm{ch1}\mathcal{A}^*_\mathrm{ch2})|\alpha_\mathrm{ch1}\alpha_\mathrm{ch2}|^2 \, .
\end{equation}
The other second-order term gives
\begin{align}
&\langle\text{sig}'_{0,\mathrm{ch1}}|\text{sig}'_{2,\mathrm{ch2}}\rangle+\langle\text{sig}'_{2,\mathrm{ch2}}|\text{sig}'_{0,\mathrm{ch1}}\rangle +\langle\text{sig}'_{2,\mathrm{ch1}}|\text{sig}'_{0,\mathrm{ch2}}\rangle+\langle\text{sig}'_{0,\mathrm{ch2}}|\text{sig}'_{2,\mathrm{ch1}}\rangle = \nonumber \\
&-(\mathcal{A}_\mathrm{ch1}\mathcal{C}_\mathrm{ch2}+\mathcal{A}_\mathrm{ch2}\mathcal{C}_\mathrm{ch1}+\mathcal{A}^*_\mathrm{ch1}\mathcal{C}^*_\mathrm{ch2}+\mathcal{A}^*_\mathrm{ch2}\mathcal{C}^*_\mathrm{ch1})|\alpha_\mathrm{ch1}\alpha_\mathrm{ch2}|^2 \, .
\end{align}
Note that these results are different from the corresponding ones for a single channel (in the sense that one cannot recover the latter by simply setting $\alpha_\mathrm{ch1}=\alpha_\mathrm{ch2}$). 
 
 Since the calculations in this section serve a merely illustrative purpose, we shall refrain from going to 4th order here. It is worth noting, however, that these results are expected from the complete calculations shown in the Letter in the limit of a gating function proportional to a Dirac delta distribution in frequency.

\end{document}